\documentstyle[12pt]{iopfix}

\eqnobysec

\def\cf{cf.\ }

\def\ie{i.e.\ }

\def\wig{_{{\rm\scriptstyle w}}}
\def\PD#1#2{\frac{\partial #1}{\partial #2}}	

  \def\BGamma{{\bf \Gamma}} 
  \def\IGamma{{\it \Gamma}} 
  \def\R{{\bold R}}          
  \def\C{{\bold C}}          
  \def\hatw{\widehat}       
  \def\wt{\widetilde}       
  \def\grad{{\nabla}}       
  \def\H{{\cal H}}          
  
  \def\J{{\cal J}}          
  \def\S{{\cal S}}          

  \def\z{{\rm \bf z}}       

  \def\({{\Bigl(}}          
  \def\){{\Bigr)}}          

  \input epsf

\begin{document}

\title{Mixed Weyl Symbol Calculus and Spectral Line Shape Theory}[Mixed Weyl 
Symbol Calculus]
\author{T A Osborn\dag,  M F Kondrat'eva\dag,  G C Tabisz\dag $\,$
  and B R McQuarrie\ddag}

\address{\dag\  Department of Physics, University of Manitoba, Winnipeg, MB,
Canada, R3T 2N2 }
\address{\ddag\ Chemical Physics Theory Group, Department of
Chemistry, University of Toronto,  Toronto, ON, Canada, M5S 1A1}

\begin{abstract}
A new and computationally viable full quantum version of line shape
theory is obtained in terms of a mixed Weyl symbol calculus. The basic
ingredient in the collision--broadened line shape theory is the time dependent
dipole autocorrelation function of the radiator-perturber system.  The
observed spectral intensity is the Fourier transform of this correlation
function.  A modified form of the Wigner--Weyl isomorphism between quantum
operators and phase space functions (Weyl symbols) is introduced in order
to describe the quantum structure of this system.  This modification  uses a
partial Wigner transform in which the radiator-perturber relative motion
degrees of freedom are transformed into a phase space dependence, while
operators associated with the internal molecular degrees of freedom are kept
in their original Hilbert space form.  The result of this partial Wigner
transform is called a mixed Weyl symbol.  The star product, Moyal bracket and
asymptotic expansions native to the mixed Weyl symbol calculus are
determined.  The correlation function is represented as the phase space
integral of the product of two mixed symbols: one corresponding to the
initial configuration of the system, the other being its time evolving
dynamical value.  There are, in this approach, two semiclassical expansions
--- one associated with the perturber scattering process, the other with the
mixed symbol  star product.  These approximations are used in combination to
obtain representations of the autocorrelation that are sufficiently simple to
allow numerical calculation.  The leading $O(\hbar^0)$ approximation recovers
the standard classical path approximation for line shapes.  The higher order
$O(\hbar^1)$ corrections arise from the noncommutative
nature of the star product.  
\end{abstract}

\newpage
\markboth{Mixed Weyl Symbol Calculus and Spectral Line Shape Theory}
{Osborn, Kondrat'eva, Tabisz and McQuarrie}

\section{Introduction}

     Collision--broadened spectral line shapes carry important information on
the time-dependent dynamical and collisional processes occurring in a
radiating medium.  For a system consisting of radiators (light emitting or
absorbing species) and perturbers (atoms, molecules, ions or electrons), the
basic ingredient in spectral line shape theory is the Fourier transform of
the time autocorrelation function of the radiator dipole.  The physical
processes here involve the dynamical evolution of the radiator as it
undergoes multiple collisions with a large number of perturbers.  The
interaction between the time dependent radiator dipole moment and the
electromagnetic field induces absorption and emission.  In this paper we
develop a phase space based representation of quantum mechanics that is
suitable for determining this dynamics.  Within this formalism there are two
different semiclassical approximations.  These are employed in tandem to
obtain representations of the autocorrelation function that are sufficiently
simple to allow numerical calculation.

The method developed here is suggested by the Moyal
description\cite{Moy49} of quantum mechanics, which is based on the
Wigner--Weyl isomorphism\cite{FOL89} between Hilbert space operators and
functions (Weyl symbols) on classical phase space.  In this
isomorphism\cite{OM95}
the canonical $(\hat q_i , \hat p_i)$ operators are transformed into linear
phase space functions $(q_i,p_i)$.  The line shape problem presents one with
two distinguished degrees of freedom:  the internal molecular  coordinates
and the relative radiator--perturber separation variables.  In many
circumstances (large mass, large impact parameter, high relative velocity,
weak pairwise interaction) one expects the relative motion to be nearly
classical.   On the other hand, evolutions of the subsystems associated with
the internal coordinates are, generally, far from classical.  This
circumstance is exploited by the introduction of a partial Wigner transform
that converts the relative motion degrees of freedom into phase space
variables, but keeps operators associated with the internal coordinates in
their original Hilbert space form.  The result of this transformation we call
a mixed Weyl symbol.

Our approach to a quantum line shape theory is strongly influenced by
the well established `classical path approximation', which is used in many
areas of atomic and molecular physics.  It assumes that the perturbers move
as classical particles, that is, along definite paths determined by the
radiator--perturber interaction. The traditional justification for this
approximation arises from the notion that one may consider the motion of
perturbers in terms of packets of translational wave functions, following the
laws of classical mechanics.  Anderson\cite{And49} in setting up his
pioneering line shape theory argued, that since the typical distances of
closest approach are about 5\AA{}, an uncertainty in position of 1\AA{} leads
to no great ambiguity in the magnitude or the type of intermolecular forces
involved in the collision.  He invoked the uncertainty principle to conclude
that the corresponding uncertainty in velocity is only a few percent for most
molecular pairs.  The approximation evidently breaks down \cite{Bar58a} when
the de Broglie wavelength of the perturber is comparable to or larger than a
characteristic distance over which the intermolecular potential varies
appreciably.  An alternative way \cite{All82} of expressing
this condition is that for the profile to be adequately treated with a
classical path, the collisions of significance must have large angular
momentum.  Such a condition is reminiscent of that required to
calculate scattering cross sections classically\cite{Mott65}.  Early
derivations of the approximation employed heuristic arguments which relied on
physical insight.  The subsequent work of Baranger\cite{Bar58a} and Smith {\it
et al} \cite{Smith68a} unified the derivation of the classical path
approximation, and in particular employed statistical mechanics to justify
the representation of the correlation function as an integral over the phase
space variables of the perturber.  The phase space average procedure
\cite{Gao91} remains a cornerstone of the theory to the present.

The system we shall consider is a dilute gas in which each of the radiator
and perturber subsystems is in equilibrium and in which the binary collision
approximation is valid.   The binary collision approximation, which implies that strong collisions between radiator and
perturber are well-separated in time, is a central assumption \cite{Bur80} in
most line broadening theories. The radiator density is sufficiently low that
the radiator--radiator interaction may be ignored.   The linewidth is
dominated by collision broadening effects; Doppler and lifetime broadening
are omitted from the analysis.  The heavy radiator approximation is thereby
imposed.  The radiator and perturber subsystems are statistically
independent; that is, the state of the perturber does not depend explicitly
on the state of the radiator and vice versa.  This assumption is frequently
termed the `lack of back reaction' in the density matrix.  Consequently the
density matrix may be factored as $\hat\rho=\hat\rho^{(R)}\hat\rho^{(P)}$,
where $\hat\rho^{(R)}$ and $\hat\rho^{(P)}$ depend only on the radiator and
perturber variables, respectively \cite{Fano57}.  Finally, since interest
lies primarily in the effect of the field--radiator interaction on the
radiator, the electromagnetic field is treated classically.  For an overview
of collision--broadening, the reader is referred to the review \cite{All82}
of Allard and Kielkopf.

A traditional starting point for the development of line shape theory is the
expression for the power gained or lost from the radiation field to the
molecular many body system in making electric dipole transitions from all
initial states $i$ to all possible final states $f$. The power produced by a
single radiator interacting with the perturbers, is given by
$\sum_{if}\hbar\omega_{if}\rho_i P_{if}$ where $P_{if}$ is the Fermi Golden
Rule probability per unit time for a transition between states $i$ and $f$
having an energy difference $E_i - E_f = \hbar\omega_{if}$. The weight factor
$\rho_i$ is the initial state density.  The frequency content of this
expression is known as the line shape or spectral profile, $I(\omega)$. If
$C_N(t)$ is the $N$-perturber dipole-dipole autocorrelation function,
the spectral profile is the Fourier integral
\begin{equation}\label{eq1.0}
I(\omega) = \frac{1}{\pi} {\rm Re}\, \int_{0}^{\infty}\,d t\,
e^{i\omega t}\,C_N(t)\, . \end{equation}
Throughout the paper we employ the version of the binary collision
approximation \cite{Bar58a,Roy80}, which expresses $C_N(t)$ via the one
perturber autocorrelation function $C(t)$, \cf(\ref{eq1.2}), namely $C_N(t) =
C(t)^N$.

In this setting, the Hamiltonian required in the one perturber
autocorrelation function is
a sum of three contributions \numparts
\begin{equation} \hatw H =\hatw H_1 + \hatw H_2 + \hatw
H_{12}\, .  \label{eq1.1a} \end{equation}
The Hamiltonian $\hatw H_1$
determines the (radiator) molecular structure, and the pair of Hamiltonians
$\hatw H_2 + \hatw H_{12}$ generate different parts of the
radiator--perturber dynamics.  Let the $d_1$ dimensional internal radiator
coordinates be the Cartesian variables
$Q=(Q_{1},\ldots,Q_{d_{1}})\in\R_{Q}^{d_{1}}$ and the radiator--perturber
coordinates given by $q=(q_{1},\ldots ,q_{d_{2}})\in\R_{q}^{d_{2}}$.  The
state spaces over these two independent coordinate manifolds are
$\H_{1}=L^{2}(\R^{d_{1}})$ and $\H_{2}=L^{2}(\R^{d_{2}})$.  The full system
Hilbert space is  $\H = \H_{1} \otimes \H_{2} =
L^{2}(\R^{d_{1}}\times\R^{d_{2}})$.  For reasons of clarity of presentation and notational
convenience, we assume that the perturber may be treated as a point particle.
This means that $q$ is a three dimensional vector ($d_2=3$). The mixed Weyl
symbol calculus and line shape theory developed here are valid for all values
of $d_1$ and $d_2$.  If desired, the one perturber Hamiltonian
(\ref{eq1.1a}) may be extended to a many perturber system.

The Hamiltonians $\hatw H_1$ and $\hatw H_2$ are simple in the sense
that they are tensor products of operators on $\H_1$ and $\H_2$, namely
${\hatw H}_1 = \hatw h_1\otimes I_2$ and ${\hatw H}_2 =I_1\otimes \hatw h_2$,
where $I_i$ is the identity on $\H_i$. The Hamiltonian $\hatw h_1$ and
its associated eigenvalue problem $\hatw{h}_1 | \Phi_n \rangle = E_n | \Phi_n
\rangle$ determine the energy spectrum and wave functions of the radiator.
The wave function $| \Phi_1 \rangle  $ is the molecular groundstate.  If
$\hatw h_{2,0}$ is the perturber kinetic energy and $\hatw v_2$ is a
$Q$--independent (isotropic) part of the intermolecular interaction energy,
then $\hatw h_2 \equiv \hatw h_{2,0} + \hatw v_2$. The operator $\hatw H_{12}$
includes all the anisotropic parts of the radiator--perturber potential.
It depends on the relative orientation of the molecular axis and the vector
$q$.  This operator is a function of both $q$ and $Q$.  For example in the
case where the radiator can be treated as a rigid rotor $(d_1=3)$, then
$\hatw H_{12}$ has the Legendre polynomial expansion
\begin{equation} \hatw
H_{12} = \sum_{l=0}^\infty V^{(l)}_{12}(|q|)P_l(\hatw q\cdot \hatw Q)\,.
\label{eq1.1b} \end{equation}
\endnumparts
The multipole potentials
$V^{(l)}_{12}(|q|)$ are phenomenologically known functions.  The presence of
the $V^{(0)}_{12}$ term in (\ref{eq1.1b}) allows one to include portions of
the isotropic potential in $\hatw H_{12}$.

The dipole autocorrelation function \cite{And49,Bar58a,Roy80} generated by
the dynamics of $\hatw H$ is
\begin{equation} \label{eq1.2}
C(t) = {\rm Tr}_{\H} \left( \hatw\mu_j
U(H;t)^\dagger \hatw\mu_j U(H;t) \hatw{\rho} \right).
\end{equation}
In the formula
above, $U(H;t)\equiv \exp(-i\hatw H t/\hbar)$ denotes Schr\"odinger picture
evolution and $\hatw\mu_j$ is the $j{\rm th}$ Cartesian component of the
radiator electric dipole.  Tensors with repeated indices such as
$j\in(1,2,3)$ are summed over.  The initial $(t=0)$ thermal state of the
system is specified by the density matrix $\hatw{\rho}=
e^{-\beta\hatw{H}}/{{\rm Tr}{e^{-\beta\hatw{H}}}}$, where $\beta=(kT)^{-1}$.
The statistical structure of the radiator--perturber system is
normally assumed to be a stationary ergodic ensemble with temperature $T$.
In this case the canonical ensemble average on the right-hand side of
(\ref{eq1.2}) is equivalent to the long time average over many
radiator--perturber collisions.   An attractive feature of the correlation
function $C(t)$ is that the effects of statistics (via $\hatw \rho$) and
dynamics (via $U(H;t)$) are clearly separated.

Our approach to computing $C(t)$ is to generalize the existing Moyal
quantum mechanics so that it can predict the $\hatw H$ dynamics in
the full Hilbert space $\H$.  In order to see the possible relevance
of the Wigner transform methods to the classical path approximation, consider
in isolation the one body (perturber) problem generated by $\hatw h_2$ on
$\H_2=L^2(\R^{d_2})$.  Adapted to this setting, the key elements \cite{OM95}
of the Moyal theory are as follows.  Operators $\hatw A$ on $\H_2$ may be
represented as functions (Weyl symbols) on phase space.  The relevant phase
space, with variables $z=(q,p)$, is that induced by the manifold
$\R_q^{d_2}$, namely $T_{2}^{\ast}\equiv T^{\ast}(\R_{q}^{d_{2}}) \simeq
\R_{q}^{d_{2}} \times \R_{p}^{d_{2}}$.  We label the Wigner
transform\cite{Wig32} map  $\sigma$, i.e.  $\sigma\hatw A(z)=A\wig(z)$, where
\begin{equation}
A\wig(z)=\int_{\R^{d_2}} dx\,e^{-ip\cdot x/\hbar} \langle
q+{\scriptstyle{1\over2}}x|\hatw A |q-{\scriptstyle{1\over2}}x \rangle \,.
\label{eq1.3} \end{equation}
The Fourier transform nature of the Wigner correspondence (\ref{eq1.3})
ensures that it has an inverse, $\sigma^{-1}$.

The Heisenberg picture evolution generated by $\hatw h_2$ has two equivalent
realizations.  In Hilbert space $\H_2$, one has $\hatw A(t) \equiv
\BGamma(h_2;t)\hatw A = U(h_2,t)^\dagger \hatw A U(h_2;t)$.  In symbol
space the equivalent of $\BGamma(h_2;t)$ is denoted by
  $\IGamma(h_2;t)\equiv\sigma\BGamma(h_2;t)\sigma^{-1}$.  This latter operator
  maps symbols to symbols.  In detail, if $A\wig(z)$ is the Weyl
  symbol of $\hatw A$ at $t=0$ and
  $A(t|z)\equiv (\hatw A(t))\wig(z)$ is the corresponding dynamical value at
  time $t$, then
\numparts
\label{eq1.4}
\begin{equation}
       \IGamma(h_2;t)\,A\wig(z) = A(t|z)\,.  \label{eq1.4a} \end{equation}
Acting on
suitable symbols, $\IGamma(h_2;t)$ defines a one parameter group with the same
structure one finds with $\BGamma(h_2;t)$, namely, $\IGamma(h_2;t_1+t_2)=
\IGamma(h_2;t_2)\IGamma(h_2;t_1)$.  The equation of motion for
$A(t|z)$ follows from taking the Wigner transform of the Heisenberg equation
for $\hatw A(t)$,
\begin{equation}
\PD{}t A(t|z) = \{A(t),h_2\}_{\rm M}(z)\,.
\label{eq1.4b} \end{equation} \endnumparts
Here $\{\cdot,\cdot\}_{\rm
M}$ is the Moyal bracket, \cf (\ref{eq2.13b}), of the symbol pair $A(t)$ and
$h_2\equiv(\hatw h_2)\wig$.

Given the solution of (\ref{eq1.4b}), expectation values in $\H_2$ are
realized as integrals over $T_2^*$.  Let $\psi\in\H_2$ be a unit normalized
initial state defining a density matrix
$\hatw\rho=|\psi\rangle\langle\psi|$.  In terms of the Wigner distribution
$w_{\psi}(z)=(h)^{-d_2}(\hatw\rho)\wig(z)$, $(h=2\pi\hbar)$;
\numparts
\label{eq1.5} \begin{equation} \langle \hatw A(t)\rangle_{\psi}  =  {\rm
        Tr}_{\H_2}\hatw\rho \hatw A(t) = \int_{T_2^*} dz\,w_{\psi}(z)\,
\IGamma(h_2;t) A\wig (z)\, .  \label{eq1.5a} \end{equation}

In the circumstances where $h_2$ is $\hbar$--independent, $\IGamma(h_2;t)$
admits a small $\hbar$ asymptotic expansion
\begin{equation}
\fl \IGamma(h_2;t)A\wig
=\sum_{n=0}^{\infty}{\hbar^{2n}\over{(2n)!}}\gamma^{(2n)}(h_2;t) A\wig
=\Big[\gamma^{(0)}(h_2;t) +
{\hbar^{2}\over{2!}}\gamma^{(2)}(h_2;t)\Big] A\wig + O(\hbar^4)\,.
\label{eq1.5b}
\end{equation}
\endnumparts
Like $\IGamma(h_2;t)$, the quantities $\gamma^{(2n)}(h_2;t)$ are maps on
symbols.  In reference \cite{OM95}  explicit formulas for
$\gamma^{(2n)}(h_2;t)$ were derived as a result of a connected graph
representation for $\IGamma(h_2;t)$.  The leading term $\gamma^{(0)}(h_2;t)$
(as has long been known \cite{Ant77,Pro83,ROB87}) is determined by the $h_2$
generated classical flow.  The higher order operator coefficients, beginning
with $\gamma^{(2)}(h_2;t)$, have the form of partial differential operators
[\cf(\ref{eq3.8c})] acting on symbols.  Combining expansion (\ref{eq1.5b}) and
(\ref{eq1.5a}) yields a semiclassical expansion for the expectation value
$\langle \hatw A(t)\rangle_{\psi}$.  With suitable restrictions on $A\wig$,
rigorous error bound estimates are available for the asymptotic expansion
(\ref{eq1.5b}).  The early work of Antonets \cite{Ant77} verified, for finite
time displacements $t\in[0,T]$, that $\IGamma(h_2,t)A\wig \rightarrow
\gamma^{(0)}(h_2;t) A\wig$ in an appropriate norm as $\hbar\rightarrow 0$.
More recently new proofs \cite{BGP98} have been constructed that obtain error
estimates that hold for arbitrary time displacements.

The principal goal of this paper is to derive a modified version of Moyal
quantum mechanics that treats the perturber degrees of freedom with a
phase space formalism like (1.5)--(1.6)
while maintaining a
consistent $\H_1$--operator valued description of the radiator degrees of
freedom.   Within this hybrid operator--symbol formalism, the Moyal bracket
induces a natural semiclassical expansion.  The leading
approximation is shown to recover the classical path approximation. Higher
order corrections are well defined and sufficiently simple to allow numerical
calculation.

\section{Mixed Weyl Symbol Calculus}

The {\it mixed Weyl symbol} is a parametric family of operators that arises
when a phase space function is quantized in a subset of its canonical
variables.  The conventional symbol is a complex valued function in phase
space, whereas the mixed Weyl symbol is operator valued.  In this section, we
modify the Wigner transform\cite{Wig32} and Weyl quantization\cite{Wey27}
procedures so that they define a mixed Weyl symbol formalism.  Within this
formalism we construct representations of the noncommutative star
product, the Moyal bracket\cite{Moy49} and the trace formulas that
determine expectation values.  The asymptotic expansions,
which form the basis of a semiclassical analysis, are also obtained.

Each of the $\H_i$ subspace components of $\H = \H_{1} \otimes \H_{2}$
has its own set of canonical operators and coordinates. We distinguish
these variables by employing upper case letters for $\H_{1}$ and lower case
for $\H_{2}$. The $d_{1}$ system classical phase space is $T_{1}^{\ast}\equiv
T^{\ast}(\R_{Q}^{d_{1}}) \simeq \R_{Q}^{d_{1}} \times \R_{P}^{d_{1}}$ with
coordinates
$Z=(Z_{1},\ldots,Z_{2d_{1}})=(Q_{1},\ldots,Q_{d_{1}};P_{1},\ldots,P_{d_{1}})$.
In the Hilbert space $\H_{1}$ over $\R^{d_{1}}$, the quantized
coordinate operators are $\hatw{Z} = (\hatw{Q},\hatw{P})=(\hatw{Q}_{1},\ldots,
\hatw{Q}_{d_{1}};\hatw{P}_{1},\ldots ,\hatw{P}_{d_{1}})$. Acting on
$\R_{Q}^{d_{1}}$ space wave functions $\Phi \in \H_{1}$, the $\hatw{Q}_{j}$ are the
operators of multiplication by $Q_{j}$ and the conjugate momentum are
$\hatw{P}_{j} = -i\hbar\partial / \partial Q_{j}$.  Likewise the $d_2$
system has phase space $T_{2}^{\ast}\equiv T^{\ast}(\R_{q}^{d_{2}})$ with
variables $z=(q_{1},\ldots ,q_{d_{2}};p_{1},\ldots,p_{d_{2}})$ and
canonical operators
$\hatw{z}=(\hatw{q}_{1},\ldots ,\hatw{q}_{d_{2}};\hatw{p}_{1},
\ldots,\hatw{p}_{d_{2}})$.

The commutation relations
\numparts 
\label{eq2.1}
\begin{equation}
[\hatw{Z}_{\alpha},\hatw{Z}_{\beta}] = i\hbar J_{\alpha\beta}^{(1)}\,, \quad
[\hatw{z}_{\mu},\hatw{z}_{\nu}] = i\hbar J_{\mu\nu}^{(2)}\,, \quad
[\hatw{Z}_{\alpha},\hatw{z}_{\mu}] = 0 \; ,
\label{eq2.1a}
\end{equation}
state the separate canonical character and mutual independence of $\hatw{Z}$ and
$\hatw{z}$. The matrices $J^{(1)}$ and $J^{(2)}$
are the standard symplectic matrices that arise on $T_{1}^{\ast}$ and
$T_{2}^{\ast}$. In block form
\begin{equation}
J^{(1)} = \left[ \begin{tabular}{cc} $0$ &  $\delta^{d_{1}}$ \\
                    $-\delta^{d_{1}}$ & $0$ \end{tabular} \right] \; , \quad
J^{(2)} = \left[ \begin{tabular}{cc} $0$ &  $\delta^{d_{2}}$ \\
                    $-\delta^{d_{2}}$ & $0$ \end{tabular} \right] \; ,
\label{eq2.1b}
\end{equation}
\endnumparts
where $\delta^{d_{i}}$ are the $d_{i}$--dimensional identity matrices. The
transformations $J$ are invertible with  $J^{-1}=J^{\rm T}=-J$.

\subsection{Partial Quantization}

In order to help formulate a statement of partial quantization, we recall the
definition of the Weyl symbol appropriate for the full phase space
$T_{1+2}^{\ast}\equiv T^{\ast} (\R_{Q}^{d_{1}}\times\R_{q}^{d_{2}}) \simeq
T_{1}^{\ast}\times  T_{2}^{\ast}$.  All the phase spaces $T_1^*$,\,$T_2^*$
and $T_{1+2}^*$ are Euclidean.  In this circumstance, the conventional Wigner
transform \cite{Wig32,OW81,FOL89} is well defined. It maps an operator
$\hatw{A}$ on $\H$ into a complex valued function that is supported on
$T_{1+2}^{\ast}$. Specifically, if $\langle X,x |\hatw{A}| X',x' \rangle$
denotes the coordinate space Dirac kernel of $\hatw{A}$, then the Weyl symbol
is
\numparts
\label{eq2.2} \begin{equation}
\fl A_{\rm w}(Z;z)\equiv \int\!\!\! \int dX \; dx \; e^{-i(P\cdot X+p\cdot
x)/\hbar} \langle
Q+{\scriptstyle{1\over2}}X,q+{\scriptstyle{1\over2}}x|\hatw{A}|Q-{\scriptstyle{1\over2}}X,q-{\scriptstyle{1\over2}}x
\rangle \, .
\label{eq2.2a}
\end{equation}
Formula (\ref{eq2.2a}) is a restatement of (\ref{eq1.3})
adjusted to the larger phase space $T_{1+2}^{\ast}$.  This (full) Wigner
transform mapping $\hatw A\mapsto A\wig$ is always denoted by $\sigma$,
independent of which Hilbert space $\H, \H_1$ or $\H_2$ that $\hatw A$ acts
on, \ie $A\wig=\sigma\hatw A$.

The invertibility of the Fourier transform in
(\ref{eq2.2a}) means that $\sigma$ has an inverse $\sigma^{-1}$. The map
$\sigma^{-1} A_{\rm w}=\hatw{A}$ is Weyl quantization. The action of
$\sigma^{-1}$ on exponential functions is $\sigma^{-1}[\exp(i U\cdot Z + i
u\cdot z)] = \exp(i U\cdot \hatw{Z} + i u\cdot \hatw{z})$, and may be viewed
as the basis of Weyl quantization. Superposition of this result provides the
quantization for arbitrary phase space functions. Suppose $a$ is the Fourier
transform of $A_{\rm w}:T_{1+2}^{\ast}\rightarrow \C$, then
\begin{eqnarray} A_{\rm
w}(Z;z)= \int\!\!\!\int dU\; du\; a(U;u)e^{i(U\cdot Z+u\cdot z)}\; ,
\label{eq2.2b} \\ \hatw{A}=\int\!\!\!\int dU\; du\; a(U;u)
e^{i(U\cdot\hatw{Z}+u\cdot\hatw{z})}\; .
\label{eq2.2c}
\end{eqnarray}
\endnumparts 
%
%
To verify that Weyl quantization (\ref{eq2.2c}) is the inverse of the
Wigner transform (\ref{eq2.2a}), it suffices to take the $|X,x\rangle$ Dirac
matrix elements of (\ref{eq2.2c}). Fourier transform identities lead one to
recover the Wigner transform (\ref{eq2.2a}).

It is evident from the structure of (\ref{eq2.2c}) and the independence of the
operators $\hatw{Z},\hatw{z}$ that one may partially quantize $A_{\rm w}$.
In this context the two relevant choices are
\numparts 
\label{eq2.3}
\begin{eqnarray}
\hatw{A}_{\rm w1}(Z) = (\tilde{\sigma}_{2}^{-1} A_{\rm w})(Z) & \equiv &
\int\!\!\!\int dU\; du\; a(U;u) e^{iU\cdot Z}e^{iu\cdot \hatw{z}}\; ,
\label{eq2.3a} \\
\hatw{A}_{\rm w2}(z) = (\tilde{\sigma}_{1}^{-1} A_{\rm w})(z) & \equiv &
\int\!\!\!\int dU\; du\; a(U;u) e^{iU\cdot \hatw{Z}}e^{iu\cdot z}\; .
\label{eq2.3b}
\end{eqnarray}
\endnumparts 
In the notation above, the maps $\tilde{\sigma}_{2}^{-1}$ and
$\tilde{\sigma}_{1}^{-1}$ are the partial Weyl quantizations of $A_{\rm w}$
with respect to the $z$ and $Z$ variables.  The object
$\hatw{A}_{\rm w1}(Z)$ is an operator family on $\H_{2}$ with a parametric
dependence on $Z$. Likewise, $\hatw{A}_{\rm w2}(z)$ is an
operator on $\H_{1}$.

The subscript labeling ${\rm w1}$ and ${\rm w2}$ used in (2.3)
is suggested by the Wigner transform point of view. For example, the
${\rm w1}$ subscript on $\hatw{A}_{\rm w1}(Z)$ reminds one that the first
argument of $\hatw{A}$, namely $\hatw{Z}$, has been dequantized to become a
parametric dependence on $Z$.

To pass from $\hatw{A}_{\rm w1}$ or $\hatw{A}_{\rm w2}$ to $\hatw{A}$,
one Weyl quantizes, respectively,  either $e^{iU\cdot Z}$ or $e^{iu\cdot z}$
in (\ref{eq2.3}). After a Fourier transform this gives \numparts
\label{eq2.4}
\begin{eqnarray}
\hatw{A}=\sigma_{1}^{-1}\hatw{A}_{\rm w1} & \equiv & \frac{1}{(2\pi)^{2d_{1}}}
\int\!\!\!\int dU\; dZ\; \hatw{A}_{\rm w1}(Z) e^{iU\cdot(\hatw{Z}-Z)}\; ,
\label{eq2.4a} \\
\hatw{A}=\sigma_{2}^{-1}\hatw{A}_{\rm w2} & \equiv & \frac{1}{(2\pi)^{2d_{2}}}
\int\!\!\!\int du\; dz\; \hatw{A}_{\rm w2}(z) e^{iu\cdot(\hatw{z}-z)}\; .
\label{eq2.4b}
\end{eqnarray}
\endnumparts 
One may implement the transform $A_{\rm w}\mapsto\hatw{A}$ via the a or
b combinations of (2.3) and (2.4).
 This freedom to choose
the order of the partial quantizations is a consequence of the mutual
commutivity of $\hatw{Z}$ and $\hatw{z}$. The product mappings that
characterize these two equivalent quantization orderings are
\numparts 
\label{eq2.5}
\begin{equation}
\hatw{A}=\sigma^{-1}A_{\rm w}=\sigma_{1}^{-1}\tilde{\sigma}_{2}^{-1}A_{\rm w}=
\sigma_{2}^{-1}\tilde{\sigma}_{1}^{-1}A_{\rm w}\; .
\label{eq2.5a}
\end{equation}

To complete the list of transformations we require the partial Wigner
transforms $\sigma_{1}\hatw{A}=\hatw{A}_{\rm w1}$ and
$\sigma_{2}\hatw{A}=\hatw{A}_{\rm w2}$. In view of (\ref{eq2.2a}), (2.3)
we have the definitions
\begin{equation}
\sigma_{1}\hatw{A}\equiv\tilde{\sigma}_{2}^{-1}\sigma\hatw{A} \; , \quad
\sigma_{2}\hatw{A}\equiv\tilde{\sigma}_{1}^{-1}\sigma\hatw{A} \; .
\label{eq2.5b}
\end{equation}
\endnumparts 
It is straightforward to verify that $\sigma_{j}^{-1}$ and $\sigma_{j}$
($j=1,2$) are mutual inverses. The reason we have elected to define
$\sigma_{1}$ and $\sigma_{2}$ by the product of transformations in
(\ref{eq2.5b}) rather than by the direct route from $\hatw{A}$ to
$\hatw{A}_{\rm w1}$ or $\hatw{A}_{\rm w2}$ via an integral like
(\ref{eq2.2a}) is that the latter requires a non--standard form of the Dirac
ket.  Namely, instead of matrix elements of $\hatw{A}$ with respect to the
full ket $|X,x\rangle$ one would need matrix elements in terms of a partial
ket $|x\rangle$.

\begin{figure}
\centerline{\hbox{\epsfxsize = 3.4 in \epsfbox{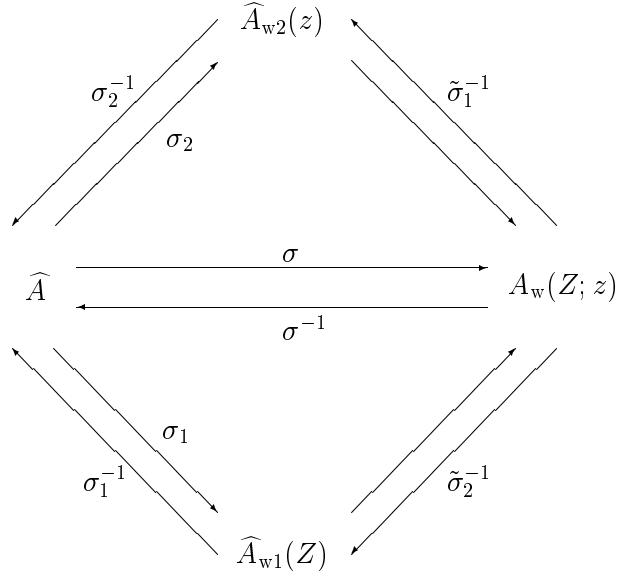}}}
\caption{Mixed Symbol Mappings}
\label{fig1}
\end{figure}

The identification and interdependence of the various mixed symbols and
partial quantizations are displayed in Fig. \ref{fig1}.

It is instructive to contrast the basic features of the full Wigner--Weyl
isomorphism $\sigma$ with those induced by the partial transformations
$\sigma_i$.  Consider the tensor product operator
$\hatw{A}=\hat{f}\otimes\hat{g}$ on $\H =\H_1\otimes\H_2$. The $\sigma_{2}$
transform of $\hatw{A}$ has the factored form
$\sigma_{2}(\hat{f}\otimes\hat{g})(z)=g(z)\hat{f}$, where
$g(z)=(\sigma\hat{g})(z)$, \cf (1.3). In this example the $z$ parametric
dependence resides in the $\C$--valued multiplier $g(z)$. This means that
the resultant families  of operators commute, e.g.
$[g(z)\hat{f},g(z')\hat{f}]=0$ for all $z,z'$. However,
for a general operator $\hatw{A}$, the different operators in the family
$\hatw{A}_{\rm w2}(z)$ will not commute.

A similar example concerns the  $\sigma_{2}$ image of
$\hat{f}\otimes\hatw{z}_{\alpha}$. Let $\pi_{\alpha}$ be the linear coordinate
functions on $T_{2}^{\ast}$, i.e. $\pi_{\alpha}(z)=z_{\alpha}$.
The $\sigma_2$ Wigner transform maps this operator into
\begin{equation}
\sigma_{2}(\hat{f}\otimes\hatw{z}_{\alpha})(z)=(\sigma\hatw{z}_{\alpha})(z)
\hat{f}=\pi_{\alpha}(z)\hat{f} \; .
\label{eq2.6}
\end{equation}

All the transformations $\sigma$, $\sigma_{1}$, $\sigma_{2}$  are linear
bijective correspondences. A simplifying aspect of these transformations is
the reality feature. Selfadjoint operators
$\hatw{A}=\hatw{A}^{\dagger}$ have real valued Weyl symbols
$A\wig(Z;z)=A\wig(Z;z)^*$, and Hermitian valued mixed symbols $\hatw{A}_{\rm
w2}(z)=\hatw{A}_{\rm w2}(z)^{\dag}$.

 \subsection{Symbol Products and the Moyal Bracket}

The $\ast$ product of symbols \cite{FOL89,BS91,Bay+78a}
is constructed so that the noncommutative product structure
of operators on Hilbert space is accurately mirrored in the
space of symbols. For operators $\hatw{X}$, $\hatw{Y}$ on $\H$ with symbols
$X_{\rm w}$, $Y_{\rm w}$ the conventional $\ast$ product is
$X_{\rm w}\ast Y_{\rm w}\equiv\sigma(\hatw{X}\hatw{Y})$. By definition,
this product is noncommutative.  The mixed symbol `star' is obtained via
the same procedure. In terms of $\hatw{X}_{\rm w2}$ and
$\hatw{Y}_{\rm w2}$, let
\numparts 
\label{eq2.9}
\begin{equation}
\hatw{X}_{\rm w2}\ast\hatw{Y}_{\rm w2}\equiv\sigma_{2}(\hatw{X}\hatw{Y})\; .
\label{eq2.9a}
\end{equation}
Of course, there is a different `star' product for each of the symbols
$A_{\rm w}$, $\hatw{A}_{\rm w1}$ and $\hatw{A}_{\rm w2}$. We denote
all of these products by the same character $\ast$. The values of the
surrounding symbols will determine the selection of the relevant `star'
operation.

A basic representation of the $\sigma_2$--star product is given by the
integral formula\cite{Poo66}
\begin{equation}
\fl\hatw{X}_{\rm w2}\ast\hatw{Y}_{\rm w2}(z)
=\left(\frac{\lambda}{\pi}\right)^{2d_{2}}
\int\!\!\!\int
dz' \; dz'' \; \hatw{X}_{\rm w2}(z+z')\hatw{Y}_{\rm w2}(z+z'')
\exp\left[{2i}{\lambda}(z'\cdot J^{(2)}z'')\right] \; ,  \label{eq2.9b}
\end{equation}
where $\lambda={\hbar}^{-1}$. Whenever the integral exists, it provides
an exact formula for $\hatw{X}_{\rm w2}\ast\hatw{Y}_{\rm w2}$.  A distinguishing feature here is that the
integrand is a product of two noncommuting $\H_{1}$ valued operators. This
formula is verified by expressing $\hatw{X}\hatw{Y}$ in terms of two copies
of (\ref{eq2.4b}) and using the Baker--Cambell--Hausdorff identity
$\exp(iu'\cdot\hatw{z})\exp(iu''\cdot\hatw{z})=
\exp(-i\frac{\hbar}{2}u'\cdot J^{(2)}u'')\exp(i(u'+u'')\cdot\hatw{z})$.
Acting with $\sigma_{2}$ on the $\hatw{X}\hatw{Y}$ product, while employing
$\sigma_{2}\exp(i(u'+u'')\cdot\hatw{z})=\exp(i(u'+u'')\cdot z)$, leads to
(\ref{eq2.9b}).  Formula (\ref{eq2.9b}) also represents the $\sigma$--star
product if the quantities $\hatw{X}_{\rm w2},\hatw{Y}_{\rm w2}$ are replaced
with $\C$--valued symbols.

The mixed symbol Moyal bracket is the $\sigma_{2}$ image of the commutator
$[\hatw{X},\hatw{Y}]$, specifically
\begin{equation}
i\hbar\{\hatw X_{\rm w2},\hatw Y_{\rm w2}\}_{\rm M}\equiv
\hatw{X}_{\rm w2}\ast\hatw{Y}_{\rm w2} - \hatw{Y}_{\rm w2}\ast\hatw{X}_{\rm w2}
=\sigma_{2}[\hatw{X},\hatw{Y} ] \; .
\label{eq2.9c}
\end{equation} \endnumparts 
As with the $\ast$ notation, the appropriate meaning of the bracket
$\{\cdot,\cdot\}_{\rm M}$ is selected by the values of its argument
symbols. In all cases, the Moyal bracket is bilinear, skew and obeys the
Jacobi identity.  The Lie algebra defined by the Moyal bracket is isomorphic
to the Lie algebra induced by the commutator of operators on $\H$.

\subsection{Star Product Asymptotics}

In many circumstances the $*$ product is close to commutative
multiplication, and the difference of these two products is described by an
asymptotic expansion involving derivatives of symbols.  In this subsection we
construct derivative expansions for mixed Weyl symbols and summarize their
asymptotic structure.

Asymptotic expansions of the $*$ product are a consequence of the
large $\lambda$ limit of the integral (\ref{eq2.9b}).  In these expansions it
is desirable to have an effective small $\hbar$ scaling, but our line shape
application requires that Planck's constant be fixed at its physical value,
$\hbar=1.055\times 10^{-34}$J$\cdot$s.
In order to accommodate these opposing demands, we set
$\lambda = (\epsilon\hbar)^{-1}$ in (\ref{eq2.9b}), and effect the small
$\hbar$ scaling by letting the dimensionless parameter $\epsilon\rightarrow
0$.

First recall Groenewold's \cite{Gro46,Vor78} expansion for $\C$--valued
symbols.  Suppose the operators $\hat f$, $\hat g$ on $\H_2$ have symbols
$f=\sigma(\hat f)$, $g=\sigma(\hat g)$.  For small $\epsilon$, one has
\numparts
\label{eq2.G1}
\begin{eqnarray}
\fl f*g(z)  =  \sum_{n=0}^{N-1}
\frac{1}{n!}\left(\frac{i\epsilon\hbar}{2}\right)^n
J^{(2)}_{\mu_{1}\nu_{1}}\cdots J^{(2)}_{\mu_{n}\nu_{n}}
f_{;\mu_{1}\cdots\mu_{n}}(z) g_{;\nu_{1}\cdots\nu_{n}}(z) +R_N(z)\; 
\label{eq2.G1a} \\
\lo=  f(z)g(z) + \frac{i\epsilon\hbar}{2}\{f,g\}(z) + O(\epsilon^2) \; .
\label{eq2.G1b}
\end{eqnarray}
\endnumparts
The first term on the right of (\ref{eq2.G1b}) is ordinary multiplication,
while the term linear in $\epsilon$ is the Poisson bracket, $\{f,g\} =
\grad f\!\cdot\! J^{(2)}\grad g$. The tensor indices $\mu_{1}\cdots\mu_{n}$
on $f$ and $g$ denote the partial derivative $\partial^{n} / \partial
z_{\mu_{1}}\cdots \partial z_{\mu_{n}}$.  The remainder $R_N(z)$ is
$O(\epsilon^N)$.

We require expansions analogous to (2.8)
in which $f$ and $g$ are
replaced by mixed Weyl symbols.  A useful derivative notation adapted to the
Groenewold type expansion is the following.  Denote by $B$ the extended
Poisson bracket operator for $T_{2}^{\ast}$.  The $B$ operator acts on a tuple
of mixed Weyl symbols $\prec\hatw{X}_{\rm w2},\hatw{Y}_{\rm w2}\succ$ to
produce a new mixed Weyl symbol.  The first and higher iterates of $B$ are
\numparts 
\begin{eqnarray}
\fl B\prec\hatw{X}_{\rm w2},\hatw{Y}_{\rm w2}\succ(z)  \equiv 
J^{(2)}_{\mu\nu}\hatw{X}_{{\rm w2};\mu}(z)\hatw{Y}_{{\rm w2};\nu}(z) \; ,
\label{eq2.11a} \\
\fl B^{n}\prec\hatw{X}_{\rm w2},\hatw{Y}_{\rm w2}\succ(z) \equiv 
J^{(2)}_{\mu_{1}\nu_{1}}\cdots J^{(2)}_{\mu_{n}\nu_{n}}
\hatw{X}_{{\rm w2};\mu_{1}\cdots\mu_{n}}(z)
\hatw{Y}_{{\rm w2};\nu_{1}\cdots\nu_{n}}(z) \; 
\label{eq2.11b} \\
\lo=  (-1)^{n} J^{(2)}_{\nu_{1}\mu_{1}}\cdots J^{(2)}_{\nu_{n}\mu_{n}}
\hatw{X}_{{\rm w2};\mu_{1}\cdots\mu_{n}}(z)
\hatw{Y}_{{\rm w2};\nu_{1}\cdots\nu_{n}}(z) \; .
\label{eq2.11c}
\end{eqnarray}
\endnumparts
The equivalent pair of representations (\ref{eq2.11b}) and (\ref{eq2.11c})
for $B^{n}$ follow from the skew nature of $J^{(2)}$, namely
$J^{(2)}_{\mu\nu}=-J^{(2)}_{\nu\mu}$. Clearly
$B\prec\hatw{X}_{\rm w2},\hatw{Y}_{\rm w2}\succ(z)$ has the derivative
structure of a Poisson bracket on $T_2^*$, but with operator valued rather
than scalar valued arguments.  The quantity $\hbar B$ is dimensionless.

The generalization of the Groenewold expansion to the mixed symbol $*$ product
is
\numparts
\label{eq2.12}
\begin{eqnarray}
\fl \hatw{X}_{\rm w2}\ast\hatw{Y}_{\rm w2}(z) = 
\exp(i\epsilon\hbar B/2)\prec \hatw{X}_{\rm w2},\hatw{Y}_{\rm w2} \succ(z)
\label{eq2.12a} \\
\lo=  \hatw{X}_{\rm w2}(z)\hatw{Y}_{\rm w2}(z) + \frac{i\epsilon\hbar}{2}
B\prec\hatw{X}_{\rm w2},\hatw{Y}_{\rm w2}\succ(z) + \cdots
 \; .
\label{eq2.12b} \end{eqnarray} \endnumparts 
Formula (2.10)
indicates how the $\ast$ operator modifies the $\H_{1}$ product of two mixed
symbols, $\hatw{X}_{\rm w2}(z)$ and $\hatw{Y}_{\rm w2}(z)$. Clearly, in the
algebra of mixed symbols, $\hatw{X}_{\rm w2}(z)$, one has two
noncommutative mechanisms --- the $\ast$ operation and the
noncommutative $\H_{1}$ operator product.

As $\lambda\rightarrow \infty$ the exponential in (\ref{eq2.9b}) oscillates
rapidly.  This circumstance justifies an application of the stationary phase
method \cite{MF81} and leads to the asymptotic expansion (\ref{eq2.12b}).
In order to establish the equivalent formal identity (\ref{eq2.12a}),
express the $\hatw{X}\hatw{Y}$ product as a multiple integral by using
representation (\ref{eq2.4b}). In the integrand one encounters
$\exp(-\frac{i\hbar}{2}u\cdot J^{(2)}u')\exp(i(u+u')\cdot z)$.  Rewrite this
as $\exp(\frac{i\hbar}{2}\nabla_{z}\cdot J^{(2)}\nabla_{z'}) \exp(iu\cdot
z+iu'\cdot z')|_{z=z'}$ and take the left exponential operator outside the
$du\,du'$ integration. This gives (\ref{eq2.12a}).

An important application of (2.10)
arises when the expansion is
used to describe the $\sigma_{2}$--Moyal bracket. Using the equivalence
of (\ref{eq2.11b}) and (\ref{eq2.11c}) the exponential series reorganizes as
\numparts 
\label{eq2.13}
\begin{eqnarray}
i\hbar \{ \hatw{X}_{\rm w2},\hatw{Y}_{\rm w2} \}_{\rm M}(z)& = &
\cos(\frac{\epsilon\hbar}{2}B)\prec\hatw{X}_{\rm w2},
\hatw{Y}_{\rm w2}\succ_{-}(z)\nonumber \\
 & + & i\sin(\frac{\epsilon\hbar}{2}B)\prec\hatw{X}_{\rm w2},\hatw{Y}_{\rm
w2}\succ_{+}(z) \;  \label{eq2.13a} \\
 = [\hatw{X}_{\rm w2}(z),\hatw{Y}_{\rm w2}(z)] &+ &\frac{i\epsilon\hbar}{2}B
\prec\hatw{X}_{\rm w2},\hatw{Y}_{\rm w2}\succ_{+}(z)  + O(\epsilon^2) \; .
\label{eq2.13b} \end{eqnarray} \endnumparts
Here $\prec\hatw{X}_{\rm w2},\hatw{Y}_{\rm w2}\succ_{\pm}=
\prec\hatw{X}_{\rm w2},\hatw{Y}_{\rm w2}\succ\pm\prec
\hatw{Y}_{\rm w2},\hatw{X}_{\rm w2}\succ$
denotes the symmetrized and anti-symmetrized tuple, respectively.
Whenever the operator families $\hatw{X}_{\rm w2}(z)$ and
$\hatw{Y}_{\rm w2}(z')$ commute, the cosine contribution vanishes and one
recovers the familiar \cite{Moy49} sine function version of the Moyal bracket.

The parameter $\epsilon\in[0,1]$ measures the deformation of the $*$ product
from the conventional $\epsilon=0$ product.  It provides a
one dimensional variable that allows one to characterize the derivative
expansions, (2.8) and (2.10),
as $\epsilon\rightarrow 0$
asymptotic approximations of the $*$ operation.  However, even if
$\epsilon = 1$ and there is no exposed small parameter, the Groenewold
formulas (2.8) and (2.10)
may still be valid asymptotic expansions.  (See Appendix A).

 \subsection{Trace Identities}

Quantum expectation values are determined, within the density matrix
formalism, by traces of operators on $\H$. For the $\sigma$--transform the
quantum trace is realized by a phase space integral. Assuming that
$\hatw{Y}$ and $\hatw{Y}\hatw{X}$ are trace class on $\H$ with smooth
symbols, the known
\cite[Chap. IV] {Shu87} trace representations are \numparts
\label{eq2.14} \begin{eqnarray}
{\rm Tr}_{\H}\hatw{Y} &=&
\frac{1}{h^{d_{1}+d_{2}}} \int\!\!\!\int_{T_{1+2}^{\ast}} dZ\; dz\; Y_{\rm
w}(Z,z) \; , \label{eq2.14a} \\ {\rm Tr}_{\H}\hatw{Y}\hatw{X} &=&
\frac{1}{h^{d_{1}+d_{2}}} \int\!\!\!\int_{T_{1+2}^{\ast}} dZ\; dz\; Y_{\rm
w}(Z,z)X_{\rm w}(Z,z) \; .  \label{eq2.14b} \end{eqnarray} \endnumparts

Our study of dynamics will systematically employ the mixed symbols
$\hatw{Y}_{\rm w2}(z)$ and $\hatw{X}_{\rm w2}(z)$, thus we require the
$\sigma_{2}$--counterparts of (\ref{eq2.14}). One finds
\numparts 
\label{eq2.15}
\begin{eqnarray}
{\rm Tr}_{\H}\hatw{Y} &=& \frac{1}{h^{d_{2}}}
\int_{T_{2}^{\ast}} dz\; {\rm Tr}_{\H_{1}} \hatw{Y}_{\rm w2}(z) \; ,
\label{eq2.15a} \\
{\rm Tr}_{\H}\hatw{Y}\hatw{X} &=& \frac{1}{h^{d_{2}}}
\int_{T_{2}^{\ast}} dz\; {\rm Tr}_{\H_{1}}\hatw{Y}_{\rm w2}(z)*
\hatw{X}_{\rm w2}(z) \; .
\label{eq2.15b}
\end{eqnarray}
\endnumparts 

The proofs of (2.13)
are simple. To show (\ref{eq2.15a}), write the
trace on $\H_{1}$ in terms of the Dirac integral kernel, viz.
${\rm Tr}_{\H_{1}}\hatw{Y}_{\rm w2}(z)=\int dX\;
\langle X|\hatw{Y}_{\rm w2}(z)|X\rangle$. Now express $\hatw{Y}_{\rm w2}(z)$
via (\ref{eq2.3b}) and use the fact that
$\langle X|\exp(iU_{1}\cdot\hatw{Q}+iU_{2}\cdot\hatw{P})|X\rangle=
\exp(iU_{1}\cdot X)\delta^{d_{1}}(\hbar U_{2})$. If one computes the
right hand side of (\ref{eq2.15a}) the various Fourier integrals all
become delta functions, giving a result that is the $dZ\,dz$ integral on the
right of (\ref{eq2.14a}). This establishes (\ref{eq2.15a}).

Formula (\ref{eq2.15b}) is a consequence of (\ref{eq2.15a}),
if $\hatw{X}$ is replaced by $\hatw{Y}\hatw{X}$ in
(\ref{eq2.15a}). A further simplification of (\ref{eq2.15b})
allows one to replace the integrand by ${\rm Tr}_{\H_{1}}
\hatw{Y}_{\rm w2}(z)\hatw{X}_{\rm w2}(z)$.
In order to see this notice that an integration by parts shows
\begin{equation} \int
dz\;J^{(2)}_{\mu\nu}\hatw{X}_{{\rm w2};\mu}(z)\hatw{Y}_{{\rm w2};\nu}(z)=
-\int dz\;J^{(2)}_{\mu\nu}\hatw{X}_{\rm w2}(z)\hatw{Y}_{{\rm w2};\mu\nu}(z)=0
\; .
\label{eq2.16}
\end{equation}
In the rightmost integral, the skew matrix $J^{(2)}_{\mu\nu}$ times the
$\mu\nu$ symmetric operator $\hatw{Y}_{{\rm w2};\mu\nu}(z)$ sums to zero.
A similar argument shows that the higher order $B^{n}$ terms vanish.

The mathematical analysis of this section is heuristic and formal in style.
Our intention has been to start from well known facts about the Wigner
transforms and to keep the derivations at the simplest level
consistent with obtaining all the structural formulas needed for the line
shape problem.  Nevertheless, we note that in the approximate theory given in
Section 4, the space $\H_1$ is replaced by an $N$--dimensional subspace
$\H_1^{(N)}$ and so the mixed symbols become $N\times N$ matrix functions.
For matrix valued symbols the rigorous methods and convergence estimates of
pseudodifferential operator analysis (\cf Appendix A) will be applicable with
straightforward modifications.

A precursor of the mixed Weyl symbol formalism, employing matrix valued
symbols, was used in \cite{Margo95} to obtain equations of motion for quantum
mean values in a Schr\"odinger evolution problem with a Hamiltonian
containing spin structure.  There one also finds an analog
of the classical path approach.

\section{Interaction Picture Dynamics}

The final form of line shape theory we devise utilizes the
interaction picture generated by the perturber Hamiltonian, $\hatw{H}_{2}$.
In this description an observable $\hatw{A}$ acquires a time dependent form
$\hatw{A}(\tau)=\BGamma(H_{2};\tau)\hatw{A}\equiv
U(\hatw{H}_2;\tau)^{\dagger}\hatw{A}U(\hatw{H}_2;\tau)$.
The operator $\BGamma(H_{2};\tau)$  denotes
the $\hatw{H}_{2}$ Heisenberg evolution on $\H$.  When this
interaction picture flow is stated in the $\sigma_{2}$--mixed symbol
representation, it becomes
\begin{equation}
\sigma_{2}\Big(\BGamma(H_{2};\tau)\hatw{A}\Big)(z)=\Bigl(\IGamma (H_{2};\tau)
\sigma_{2}\hatw{A}   \Bigr)   (z)   \equiv   \hatw{A}_{\rm w2}(\tau|z)\;   .
\label{eq3.1} \end{equation} Here   $\IGamma (H_{2};\tau)\equiv\sigma
_2\BGamma(H_2;\tau)\sigma^{-1}_2 $ denotes  the fundamental  evolution
operator that  maps  the initial mixed symbol $\hatw{A}_{\rm w2}(z)$  into
its dynamical value $\hatw{A}_{\rm w2}(\tau|z)$.

In this section, a semiclassical approximation of $\IGamma
(H_{2};\tau)$ based on the quantum trajectories generated by
$\hatw{H}_{2}=I_1\otimes\hatw{h}_{2}$ is constructed. The structure
of the expansion closely parallels that found in Section I  cf. (1.6)
and in references \cite{OM95,MCQ98}.  This prior work developed the
semiclassical asymptotics for the ($\C$--valued) Weyl symbol description of
quantum mechanics. However, in the present circumstance $\hatw{A}_{\rm
w2}(\tau|z)$ is an $\H_{1}$--operator valued symbol.  The approach
introduced in reference \cite{OM95} was to use quantum trajectories as a basis
from which to construct the full semiclassical expansion. The arguments below
show that this idea remains applicable to the study of the mixed Weyl symbol
evolution $\IGamma (H_{2};\tau)$.

In Section 2, the dimensionless parameter $\epsilon$ was introduced to
describe the idea of scaling $\hbar$ to zero.  We maintain this notation in
this section and display this scaling as $\epsilon\hbar\rightarrow 0$.
The semiclassical representations found in this section presume that the
$\sigma_2$ symbols for $\hatw H_2$ and $\hatw A$ admit
$\epsilon\hbar\rightarrow 0$ asymptotic expansions.  This is indeed the case
here, where both $\hatw H_2$ and $\hatw A$ have $\hbar$ independent symbols.

\subsection{Quantum Trajectories}

Let the static canonical operators $\{\hatw{z}_{\alpha}\}_{1}^{2d_{2}}$ be
restricted to the space $\H_{2}$ and let
$\hatw{z}_{\alpha}(\tau)=\BGamma(h_2;\tau)\hatw z_\alpha$ denote the
associated $\hatw {h}_{2}$ Heisenberg evolution. The quantum
trajectories are defined as ${\z}_{\alpha}(\tau |z) = \IGamma
(h_2;\tau)\pi_\alpha(z) = \big(\hatw z_\alpha(\tau)\big)\wig(z)$ and obey the
equation of motion 
\numparts
\label{eq3.3}
\begin{equation} \PD{}t{\z}_{\alpha}(\tau|z) = \{\z_{\alpha}(\tau),h_{2}
\}_{\rm M}(z)\; .  \label{eq3.3a} \end{equation}
In this identity, $\{\cdot,\cdot\}_{\rm M}$ is the Moyal bracket defined for $\C$--valued Weyl
symbols supported on $T_{2}^{\ast}$. At $\tau=0$, the initial condition for
$\z_{\alpha}$ is $\z_{\alpha}(0|z)=\pi_{\alpha}(z)=z_\alpha$.  Equation
(\ref{eq3.3a}) is an important example of (\ref{eq1.4b}).

An approximate solution of $\z_\alpha(\tau|z)$ is obtained by expanding
(\ref{eq3.3a}) in small $\epsilon$. That this is possible is
a result of the standard\cite{FOL89} small $\epsilon$ asymptotic expansion of
the Moyal bracket, \cf (2.8)
\begin{equation}
\fl \{\z_{\alpha}(\tau),h_{2}\}_{\rm M}(z)=\{\z_{\alpha}(\tau),h_{2}\}(z)+
\sum_{n=1}^{\infty}\frac{(-1)^{n}(\epsilon\hbar/2)^{2n}}{(2n+1)!}B^{2n+1}
\prec \z_{\alpha}(\tau),h_{2}\succ (z)\; .
\label{eq3.3b}
\end{equation}
Using (\ref{eq3.3b}) in (\ref{eq3.3a}) and solving for the coefficient
functions order by order in $\epsilon$, one finds\cite[Sec. III]{OM95}
\begin{equation}
\fl\z_{\alpha}(\tau|z)
=\sum_{n=0}^{\infty}\frac{(\epsilon\hbar)^n}{n!}\z_{\alpha}^{(n)}
(\tau|z)=g_{\alpha}(\tau|z)+\frac{(\epsilon\hbar)^2}{2}\z_{\alpha}^{(2)}(\tau|z)
+O(\epsilon^4)\; .
\label{eq3.3c}
\end{equation}
The leading term above is the solution of Hamilton's equation
\begin{equation}
\dot{g}(\tau|z)=J^{(2)}\nabla h_2(g(\tau|z))\; ,
\label{eq3.3d}
\end{equation}
\endnumparts 
with initial condition $g(0|z)=z$. Because both the Moyal bracket and $h_2$
are even functions of $\hbar$, one can prove that $\z_{\alpha}(\tau|z)$ (cf.
\cite{OM95} Lemma 5) is an $\hbar$-even function. As a result,
all $n$ odd terms vanish in expansion (\ref{eq3.3c}). The higher order
correction functions $\z_{\alpha}^{(n)}(\tau|z)$, $n \ge 2$, are obtained by
solving an inhomogeneous Jacobi field equation, (\cf\ref{eq3.10b}). The
classical flow $g(\tau|z)$ and coefficients $\z_{\alpha}^{(2)}(\tau|z)$ are
defined and finite for all time displacements $\tau$.

\subsection{Semiclassical Asymptotics of $\IGamma (H_{2};\tau)$}

In this subsection, the $\epsilon\hbar\rightarrow 0$ asymptotic expansion for
$\IGamma (H_{2};\tau)$ is derived. To start, we note that it is
useful to reorganize the Fourier integral formulas for $\hatw{A}$
and $\hatw{A}_{\rm w2}(z)$ in the following manner. Given $A_{\rm w}(Z;z)$
with Fourier dual $a(U;u)$, set
\numparts 
\label{eq3.4}
\begin{equation}
\hatw{A}(u)\equiv\int dU\; a(U;u)\, e^{iU\cdot\hatw{Z}}\; .
\label{eq3.4a}
\end{equation}
The quantity $\hatw{A}(u)$ is a $u$--parameter function with $\H_1$ operator
values. Representations (\ref{eq2.2c}) and (\ref{eq2.3b}) now read
\begin{eqnarray}
\hatw{A}=\int du\; \hatw{A}(u) \,e^{iu\cdot\hatw{z}} \; ,
\label{eq3.4b} \\
\hatw{A}_{\rm w2}(z)=\int du\; \hatw{A}(u)\,e^{iu\cdot z} \; .
\label{eq3.4c}
\end{eqnarray}
The action of $\BGamma (H_2;\tau)$ on $\hatw{A}$ has the integral form
\begin{equation}
\BGamma (H_2;\tau)\hatw{A} = \int du\; \hatw{A}(u)\,
e^{iu\cdot\hatw{z}(\tau)}\; .  \label{eq3.4d} \end{equation}
Note that the integrand in (\ref{eq3.4d}) may be restated as
$e^{iu\cdot\hatw{z}(\tau)}=e^{(\hatw{z}(\tau)-s)\cdot\nabla_{z}}
e^{iu\cdot(s+z)}|_{z=0}$. Here $s$ is an arbitrary vector in $\R^{d_{2}}$, and
$\hatw{z}(\tau)-s$ is shorthand for $\hatw{z}(\tau)-sI_2$. Combining
this with (\ref{eq3.4d}) gives one an operator valued power series
\begin{eqnarray}
\BGamma (H_2;\tau)\hatw{A}& = & \exp[(\hatw{z}(\tau)-s)\cdot\nabla_{z}]
\hatw{A}_{\rm w2}(s+z) \left.\right|_{z=0} \; 
\nonumber \\
& = & \sum_{n=0}^{\infty}\frac{1}{n!}(\hatw{z}(\tau)-s)_{\mu_{1}}\cdots
(\hatw{z}(\tau)-s)_{\mu_{n}}\hatw{A}_{{\rm w2};\mu_{1}\cdots\mu_{n}}(s)\; .
\label{eq3.4e}
\end{eqnarray}
\endnumparts 
Obviously, in this formula,
$\hatw{A}_{{\rm w2};\mu_{1}\cdots\mu_{n}}(s)$ is a
static operator, while all the dynamics resides in the $\hatw{z}(\tau)-s$
factors. Convert (\ref{eq3.4e}) into an identity  for
$\IGamma (H_{2};\tau)$ by the application of the transform $\sigma_2$ to obtain
\begin{equation}
\fl \IGamma (H_2;\tau)\hatw{A}_{\rm w2}(z)= \sum_{n=0}^{\infty}\frac{1}{n!}
(\z(\tau)-s)_{\mu_{1}}\ast\cdots\ast(\z(\tau)-s)_{\mu_{n}}(z)
\hatw{A}_{{\rm w2};\mu_{1}\cdots\mu_{n}}(s)\; .
\label{eq3.5}
\end{equation}

The series (\ref{eq3.5}) is an $s$--dependent family of representations for
the $s$--independent object $\IGamma (H_2;\tau)\hatw{A}_{\rm w2}(z)$. An
optimal choice of $s$ will make the $n{\rm th}$ series coefficient of order
$O(\epsilon^{l(n)})$, where $l(n)\ge n/2$.  For each given  $\tau,z$ choose
$s=\z(\tau|z)$. This makes the $n=1$ term zero, and gives us the expansion
\numparts 
\label{eq3.6} 
\begin{eqnarray} 
\fl \IGamma
(H_2;\tau)\hatw{A}_{\rm w2}(z) = \hatw{A}_{\rm w2}(\z(\tau|z)) +
\sum_{n=2}^{\infty}\frac{1}{n!}W_{\mu_{1}\cdots\mu_{n}}(\tau|z)
\hatw{A}_{{\rm w2};\mu_{1}\cdots\mu_{n}}(\z(\tau|z))\; ,
\label{eq3.6a} \\
\fl W_{\mu_{1}\cdots\mu_{n}}(\tau|z)  =
S_{n}(\z_{\mu_{1}}(\tau)-s_{\mu_1})\ast
\cdots\ast(\z_{\mu_{n}}(\tau)-s_{\mu_n})(z)\left.\right|_{s=\z(\tau|z)}\; .
\label{eq3.6b}
\end{eqnarray}
\endnumparts 
Here $S_{n}$ is the permutation group operator $(S_n^2=S_n)$ that averages
over all the $n!$ reorderings of the indices ($\mu_{1},\cdots,\mu_{n}$). The
$\ast$ products in (\ref{eq3.6b}) must be evaluated before the constraint
$s=\z(\tau|z)$ is imposed. In this evaluation the quantity $s$ is a
$z$--independent constant. The average $S_{n}$ ensures that the coefficient
functions $W$ are real and permutation invariant in the indices
($\mu_{1},\cdots,\mu_{n}$).

An efficient way to compute the $W$ functions is via the link operator $L$
that expresses the extent to which the $\ast$ product differs from
commutative multiplication. This operator is
$L_{jk}\equiv\exp( \frac{i\epsilon\hbar}{2}B_{jk} )-1$. It acts on an $n$--tuple
of phase space functions
$\prec A_{1}(z^{(1)}),A_{2}(z^{(2)}),\cdots,A_{n}(z^{(n)})\succ$. The
labels $jk$ specify which pair of the $n$--tuple elements that $B$ acts on
prior to diagonal evaluation at $z^{(1)}=\cdots=z^{(n)}=z$. The link
operator is $O(\epsilon)$. A short calculation gives
\numparts 
\label{eq3.7}
\begin{eqnarray}
\fl W_{\mu_1\mu_2}(\tau|z) = L_{12}S_{2}
\prec \z_{\mu_1}(\tau),\z_{\mu_2}(\tau)\succ(z)\; 
\nonumber \\
\lo=\frac{1}{8}(\epsilon\hbar)^2 B_{12}^{2}
S_{2}\prec \z_{\mu_1}(\tau),\z_{\mu_2}(\tau)\succ(z) + O(\epsilon^4) \; ,
\label{eq3.7a} \\
\fl W_{\mu_1\mu_2\mu_3}(\tau|z) = (L_{12}L_{13}L_{23}-L_{12}L_{23})S_{3}
\prec \z_{\mu_1}(\tau),\z_{\mu_2}(\tau),\z_{\mu_3}(\tau) \succ(z) \; 
\nonumber \\
\lo= \frac{1}{4}(\epsilon\hbar)^2 B_{12}B_{23}S_{3}
\prec \z_{\mu_1}(\tau),\z_{\mu_2}(\tau),\z_{\mu_3}(\tau) \succ(z)
+O(\epsilon^4)\; .
\label{eq3.7b}
\end{eqnarray}
\endnumparts 

A couple of remarks about representation (3.5)--(3.6)
are in order. If $\hatw{A}_{\rm w2}(z)$ is a polynomial in $z$, then the series
(\ref{eq3.6a}) truncates at finite order and provides exact
expressions for $\IGamma (H_2;\tau)\hatw{A}_{\rm w2}(z)$.  In essence
expansion (3.5)
is a semiclassical expansion for an arbitrary mixed
symbol $\hatw A_{\rm w2}$. It employs $\z(\tau|z)$ (quantum flow)
transport with higher order corrections arising from
$W_{\mu_{1}\cdots\mu_{n}}(\tau|z)$.  The link operator $L$ was introduced in
reference \cite{OM95} in order to determine the symbol of an exponential
operator $\exp\hatw{A}$ in terms of $A_{\rm w}$.

\subsection{The Standard Approximation}

Although expansion (3.5)
is the basic semiclassical expansion for
$\IGamma (H_2;\tau)$, it is nevertheless not convenient for numerical
calculation. If $H_{2}$ is $z$ quadratic then $\z
(\tau|z)=g(\tau|z)$ and $W_{\mu_{1}\cdots\mu_{n}}(\tau|z) = 0$. In
this case the first term is exact, \ie
$\IGamma(H_2;\tau)\hatw{A}_{\rm w2}(z) = \hatw{A}_{\rm w2}(g(\tau|z))$.
Generally, for $n>0$, $\z^{(n)}(\tau|z)$ and
$W_{\mu_{1}\cdots\mu_{n}}(\tau|z)$ do not vanish and for increasing $n$ these
functions are difficult to numerically compute.

In order to build a more computationally accessible approximation one
combines expansion (\ref{eq3.3c}) for $\z (\tau|z)$ with (3.5)
and collects all terms of common $\epsilon$ order. 
This results in an expansion of the form
\numparts 
\label{eq3.8}
\begin{equation}
\IGamma (H_2;\tau)\hatw{A}_{\rm w2}(z)= \sum_{n=0}^\infty
\frac{(\epsilon\hbar)^{2n}}{(2n)!}\gamma^{(2n)}(H_2;\tau) \hatw{A}_{\rm
w2}(z)\,.  \label{eq3.8a} \end{equation}

The coefficient operators $\gamma^{(n)}(H_{2};\tau)$ are composed of
transport along $g(\tau|z)$ plus a $z$--derivative structure inherited from
(\ref{eq3.6a}). The first term is pure classical transport
\begin{equation}
\gamma^{(0)}(H_2;\tau)\hatw{A}_{\rm w2}(z)=\hatw{A}_{\rm w2}(g(\tau|z))\; ,
\label{eq3.8b}
\end{equation}
while the $O(\epsilon^2)$ term reads
\begin{eqnarray}
\fl \gamma^{(2)}(H_2;\tau)\hatw{A}_{\rm w2}(z)=\z_{\mu}^{(2)}(\tau|z)
\hatw{A}_{{\rm w2};\mu}(g(\tau|z)) -
\frac{1}{8}w_{\mu\nu}(\tau|z)\hatw{A}_{{\rm w2};\mu\nu}(g(\tau|z))
\label{eq3.8c} \\
\lo +\frac{1}{12}w_{\mu\nu\lambda}(\tau|z)
\hatw{A}_{{\rm w2};\mu\nu\lambda}(g(\tau|z))\; .
\nonumber
\end{eqnarray}
\endnumparts 

The lower case $w$ coefficients result from the suitably scaled
limits of the functions $W$.  Since $W = O(\epsilon^2)$ for $W$'s with
two and three indices, one has
\numparts
\label{eq3.9}
\begin{eqnarray} 
\fl w_{\mu\nu}(\tau|z)=\lim_{\epsilon\rightarrow
0}(\epsilon\hbar)^{-2} W_{\mu\nu}(\tau|z)=
B_{12}^{2}\prec g_{\mu}(\tau),g_{\nu}(\tau) \succ(z) \; ,
\label{eq3.9a} \\
\fl w_{\mu\nu\lambda}(\tau|z)=\lim_{\epsilon\rightarrow 0}(\epsilon\hbar)^{-2}
W_{\mu\nu\lambda}(\tau|z)
=B_{12}B_{23}\prec g_{\mu}(\tau),g_{\nu}(\tau),g_{\lambda}(\tau)\succ(z)
\nonumber \\
\lo= -J^{(2)}\nabla g_{\mu}(\tau|z)\cdot \nabla\nabla g_{\nu}(\tau|z)\cdot
J^{(2)}\nabla g_{\lambda}(\tau|z)\; .
\label{eq3.9b}
\end{eqnarray}
\endnumparts
These formulas make it evident that the $w$ coefficients are functions of the
 classical flow $g(\tau|z)$.

A common building block in the expressions for $\gamma^{(n)}(H_2;\tau)$ is
the Jacobi field  $\nabla g(\tau|z)$ and its higher order derivatives
$\nabla\nabla g(\tau|z)$, etc. Jacobi fields describe the stability of a
trajectory $g(\tau|z)$ with respect to small changes of its initial
data. Differentiating (\ref{eq3.3d}) in the parameter $z$ yields
\numparts
\label{eq3.10}
\begin{equation}
\J(\tau)\nabla g(\tau|z)\equiv\left[ \frac{d}{d\tau}-J^{(2)}\nabla\nabla h_{2}
(g(\tau|z))\right]\nabla g(\tau|z) = 0\; .
\label{eq3.10a}
\end{equation}
The solutions of (\ref{eq3.10a}) with zero right--hand side are called Jacobi
fields. The initial condition is $\nabla g(0|z)=\delta$ (the $2d_2$
identity matrix). Related functions $\nabla\nabla g(\tau|z)$ solve a
modified form of (\ref{eq3.10a}) with a non--zero inhomogenous term
\cite[Eq. (3.11)]{MCQ98}.  Finally, $\z^{(2)}(\tau|z)$ is the solution of
the $2d_2$ system of ODE's \begin{eqnarray}
\J(\tau)_{\mu\nu}\z_{\nu}^{(2)}(\tau|z)& =&
-\frac{1}{8}w_{\mu_1\mu_2}(\tau|z)
J^{(2)}_{\mu\lambda}h_{2;\mu_1\mu_2\lambda}(g(\tau|z)) \nonumber \\
&&\mbox{}+\frac{1}{12}w_{\mu_1\mu_2\mu_3}(\tau|z)J^{(2)}_{\mu\lambda}
h_{2;\mu_1\mu_2\mu_3\lambda}(g(\tau|z))\; ,
\label{eq3.10b}
\end{eqnarray}
\endnumparts
with initial conditions $\z_{\mu}^{(2)}(0|z)=0$.

We refer to expansion (3.7)
as the {\em standard semiclassical
expansion} of $\IGamma (H_2;\tau)$. The basic
structure of $\gamma^{(n)}(H_2;\tau)$ is consistent with that derived in
reference \cite{OM95}. The new feature here is that the target object
$\hatw{A}_{\rm w2}$ is operator valued.  In the case where the target symbol
is $\C$--valued, the existence of small $\epsilon$ expansions of the standard
form have been known for a considerable time. These alternate approaches
\cite{Pro83,ROB87,OM95,BGP98} are based on restructuring the Moyal equation of
motion into a classical inhomogeneous Poisson bracket equation of motion.
The inhomogenous component is formed from the non-leading terms of the
derivative expansion of the bracket $\{\cdot,\cdot\}_{\rm M}$. Again one finds the leading $O(\epsilon^0)$ term is
classical transport. However, the higher order $O(\epsilon^n), n\ge 2$, terms
have representations \cite[Eq.(4.16)]{OM95} that are substantially
more complicated than the $\gamma^{(2)}(H_2;\tau)$ formula (\ref{eq3.8c}).

Recently an extensive numerical study \cite{MCQ98} of the effectiveness of
the two term asymptotic expansion (\ref{eq1.5b}) was carried out. For systems
with identical atom--atom pairs, such as helium, neon and argon, the time
evolution of quantum expectation values was computed. The pair interaction in
these systems was a phenomenologically determined Lennard--Jones potential.
There was no difficulty in computing the functions ($g_\mu(\tau|z),
\z^{(2)}_\mu(\tau|z), w_{\mu\nu}(\tau|z)$ and $w_{\mu\nu\lambda}(\tau|z)$)
which enabled the construction of $\gamma^{(0)}(h_2;\tau)$ and
$\gamma^{(2)}(h_2;\tau)$. The scattering problem was investigated for a variety
of Gaussian initial states and observables. In most instances the
$\gamma^{(0)}(h_2;\tau)$ term dominated the contributions of
$\gamma^{(2)}(h_2;\tau)$ to the expectation value.

One factor favoring the good convergence of expansion (\ref{eq1.5b}), in
the problems of interest here, is that the classical system is
completely integrable. This means that the classical flow is not chaotic.
In particular, this implies that for almost all $z \in T^{\ast}_{2}$ the
Jacobi field $\nabla g(\tau|z)$ can not have exponential growth in $\tau$.
Nevertheless, there exists a set of unstable classical trajectories in this
problem. All these motions are associated with an unstable equilibrium point.
These points occur when the radial potential energy
$v_{e}(r)\equiv v(r)+L^2/(2mr^2)$ (for a given angular momentum $L$) has
$v'_{e}(r)=0$, with $v''_{e}(r) < 0$. Associated with this fixed point one
obtains unstable orbits with positive Lyapunov exponent. The numerical
studies we completed establish that the $n=2$ term expansion (\ref{eq1.5b})
is inaccurate for large time displacements in the small region of phase space
surrounding the unstable fixed points .  In this neighborhood of phase space,
one must devise a different approximation for $\IGamma (h_2;\tau)$.  Several
alternatives for overcoming this difficulty are offered in Section 5.

\section{Autocorrelation Representations}

The evolution $\BGamma(H;t)\hatw \mu_j = U(H;t)^{\dagger}\hatw \mu_j
U(H;t)$ controls the time behavior of the autocorrelation function, $C(t)$.
This section combines the mixed Weyl symbol formalism and the $\IGamma(H_2;t)$
semiclassical expansion of Section 3 to construct numerically
accessible approximate representations of $C(t)$.

First it is helpful to reformulate $C(t)$, \cf (\ref{eq1.2}), in terms of
the $H_2$ picture evolution. Let \numparts \label{eq4.1} $\wt U(t)
\equiv U(H;t) U(H_2;t)^{\dagger}$;  within the $\hatw H_2$ interaction
picture framework, $\wt U(t)$ determines the full $\hatw H$ dynamics.  From
the Schr\"odinger evolution equations for $U(H;t)$ and $U(H_2;t)$, it follows
that $\wt U(t)$ is generated by the time dependent Hamiltonian,
\begin{eqnarray}
 \wt H(t) \equiv \BGamma(H_2;-t)(\hatw H_1 + \hatw H_{12}) = \hatw H_1 +
\BGamma(H_2;-t)\hatw H_{12}\,, \label{eq4.1a}
\\
   i\hbar \PD{}t \wt U(t) = \wt U(t) \wt H(t)\,. \label{eq4.1c}
\end{eqnarray}
\endnumparts
The operator $\wt U(t)$ is unitary and has initial condition $\wt U(0)=I$.
The appearance of $\wt H(t)$ to the right of $\wt U(t)$ is characteristic of
a backward evolution equation.

Conjugation of $\hatw A$ by $\wt U(t)$  defines a Heisenberg
evolution $\wt \BGamma(t)$ on $\H$ in the standard way, $\wt \BGamma(t)\hatw
A \equiv \wt U(t)^{\dagger} \hatw A \wt U(t)$. Knowledge of
the evolutions $\wt \BGamma(t)$ and $\BGamma(H_2;-t)$ provides an
alternate way of determining the correlation function.  Employing the cyclic
invariance of the trace and the commutation relation $[\hatw\mu_j, \hatw
H_2]=0$ allows one to represent (\ref{eq1.2}) as
\label{eq4.2n} \begin{equation}
C(t) = {\rm Tr}_{\H} \left[\left(\wt \BGamma(t)\hatw\mu_j\right)
\left(\BGamma(H_2;-t)\hatw{\rho}\right)\hatw\mu_j\right] \,.
\end{equation}
Whenever $[\hatw\rho, \hatw H_2] = 0$, the density matrix
is time independent, \ie $\BGamma(H_2;-t)\hatw\rho = \hatw\rho$. In
most applications, the anisotropic interaction $\hatw H_{12}$ is small
relative to $\hatw H_1$ and $\hatw H_2$.  This means that the radiator and
perturber systems are weakly coupled and that the density matrix is
accurately approximated by the tensor product of the form,  $\hatw
\rho_1(h_1)\otimes\hatw \rho_2(h_2)$. For a system with radiators initially
in the ground state, we may take the effective density matrix to be
$\hatw{\rho}= |\Phi_1\rangle\langle\Phi_1 |\otimes e^{-\beta\hatw{h}_2}$.  We
use this latter form of the density matrix throughout Section 4.

\subsection{Exact Mixed Symbol Dynamics}

The trace identity (\ref{eq2.15b}) determines $C(t)$ in terms of the $T_2^*$
phase space integral of $(\wt \BGamma(t)\hatw \mu_j)_{\rm w2}(z)$ and
$(\hatw \rho \hatw \mu_j)_{\rm w2}(z) = (e^{-\beta\hatw h_2})\wig(z)
|\Phi_1\rangle\langle\Phi_1 |\hatw \mu_j$.
At this stage one needs an
equation of motion for $\wt \BGamma(t)$ dynamics which is stated
in terms of a mixed symbol evolution operator:
$(\wt \BGamma(t)\hatw A)_{\rm w2}(z) = \wt \IGamma(t)\hatw A_{\rm w2}(z)$,
where $\wt \IGamma(t)\equiv \sigma_2 \wt \BGamma(t) \sigma_2^{-1}$.
The symbol valued Heisenberg flow $\wt \IGamma(t) {\hatw A}_{\rm w2}$
obeys
\numparts\label{eq4.2} \begin{equation} \PD{}t
\wt \IGamma(t) \hatw A_{\rm w2} = \{\wt \IGamma(t) \hatw A_{\rm w2},\, {\wt
H}(t)_{\rm w2}\}_{\rm M} \label{eq4.2a} \end{equation}
with Moyal bracket (\ref{eq2.9c}).  In the present application,
$\hatw A_{\rm w2}$ is the dipole operator $\hatw \mu_j$.  The
Hamiltonian here is the mixed symbol image of (\ref{eq4.1a}),
\begin{equation} \label{eq4.2b} {\wt
H}(t)_{\rm w2}(z) = \hatw H_1 + \IGamma(H_2;-t)(\hatw H_{12})_{\rm w2}(z)\,.
\end{equation}
The matrix element version of (\ref{eq4.2a}) with respect to
the basis $\{|\Phi_k\rangle\}_{k=1}^\infty$ reads
\begin{eqnarray}
\label{eq4.2c}
i\hbar {\PD{}t} X^j_{kr}(t|z) & = & 
\sum_{s=1}^\infty \Big( X^j_{ks}(t|z)*
\langle \Phi_s|{\wt H}(t)_{\rm w2}(z)|\Phi_r\rangle
\nonumber\\
& - & 
\langle \Phi_k|{\wt H}(t)_{\rm w2}(z)|\Phi_s\rangle*
X^j_{sr}(t|z) \Big)\, .
\end{eqnarray} 
\endnumparts
where $X^j_{kr}(t|z)\equiv \langle \Phi_k|\big(\wt \IGamma(t) \hatw
\mu_j\big) (z)|\Phi_r\rangle$.  This latter definition implies the
$z$--independent initial condition
$X^j_{kr}(0|z)\equiv \langle\Phi_k|\hatw \mu_j|\Phi_r \rangle $.

Given the solution of (\ref{eq4.2c}), the correlation function is
represented as
\begin{eqnarray}\label{eq4.3} 
C(t) & = & \frac{1}{h^{3}}
\int_{T_{2}^{\ast}} dz\; {\rm Tr}_{\H_1} (\wt \IGamma(t) \hatw \mu_j)(z)*(\hatw
\rho \hatw \mu_j)_{\rm w2}(z)\, \nonumber \\ & = &
\frac{1}{h^{3}}\sum_{k=1}^\infty \int_{T_{2}^{\ast}} dz\; X^j_{k1}(t|z)
X^j_{1k}(0|z) (e^{-\beta \hatw h_2})\wig(z)\,.
\end{eqnarray}
The second version of (\ref{eq4.3}) uses the complete basis
$\{|\Phi_k\rangle\}_{k=1}^\infty$
to evaluate the $\H_1$ trace.  Since the integrand is just
the product of two symbols, \cf (\ref{eq2.16}), the $*$ operation may be
removed.  Throughout this section the
dimension $d_2=3$.

\subsection{Approximate Mixed Symbol Dynamics}

For typical molecular systems, equations (\ref{eq4.2c}) are too difficult
to solve exactly. Further progress depends on developing approximate
solutions that take advantage of the type of physics present in the line
shape problem and the opportunities for simplification inherent in the mixed
symbol formalism.  The statement of exact dynamics  provides
a point of departure for the construction of various approximating methods.
Possible options are the conversion of (\ref{eq4.2c}) into an
integral equation, the use of perturbation theory, and the development of an
eikonal representation.  The approach we take here is to focus on the
semiclassical structure.  The subsequent reductions of (4.3)
rest on three approximations:

1)\indent It is assumed that only a small number of $\hatw h_1$
eigenstates $\{|\Phi_k\rangle\}_{k=1}^N$ significantly couple to each other.  This
means that $\H_1$ is replaced by the finite dimensional vector space,
$\H_1^{(N)}$, spanned by the basis $\{|\Phi_k\rangle\}_{k=1}^N$.

 2)\indent The Hamiltonian, ${\wt H}(t)_{\rm w2}(z)$, may be approximated by
the standard semiclassical expansion of Section 3, namely,
\begin{eqnarray}\nonumber
\fl {\wt H}(t)_{\rm w2}(z) = \hatw H_1 + \sum_{n=0}^\infty
\frac{(\epsilon\hbar)^{2n}}{(2n)!}\gamma^{(2n)}(H_2;-t) (\hatw {H}_{12})_{\rm
w2}(z)\,.  \\
\lo= \hatw H_1 + (\hatw H_{12})_{\rm w2}(g(-t|z)) +
\frac{(\epsilon\hbar)^{2}}{2!}\gamma^{(2)}(H_2;-t) (\hatw {H}_{12})_{\rm w2}(z)
+ O(\epsilon^4)\,. \label{eq4.4}
\end{eqnarray}

3)\indent The Moyal bracket on the right of (\ref{eq4.2a}) can be replaced
with leading terms of the small $\epsilon$ expansion (\ref{eq2.13b}).

Solvable reduced equations of motion are realized by applying
approximations 1)--3).
The finite coupled state assumption 1) is applicable when the range of thermal
energies available in the collision process can excite a limited
set of radiator eigenstates.  The next stage is to employ the
semiclassical expansion 2) valid to order $O(\epsilon^2)$.  In this
approximation, the $H_{12}$ part of ${\wt H}_{\rm w2}(t)$ defines an
$O(\epsilon^1)$ consistent time and state dependent molecular interaction by
$M_{kr}(t|z) \equiv \langle \Phi_k|({\hatw H}_{12})_{\rm
w2}\big(g(-t|z)\big)| \Phi_r\rangle$.  The modified version of (\ref{eq4.2c})
thus becomes the $N\times N$ system 
\begin{eqnarray} 
i\hbar {\PD{}t}
\chi^j_{kr}(t|z) &=& (E_r - E_k)\chi^j_{kr}(t|z) \nonumber\\
&+& \sum_{s=1}^N
\bigg[\Big(\chi^j_{ks}(t|z)*M_{sr}(t|z) - M_{ks}(t|z)*\chi^j_{sr}(t|z)\Big)
\bigg]\,.  \label{eq4.5} 
\end{eqnarray} 
The initial condition for
(\ref{eq4.5}) is $\chi^j_{kr}(0|z)=\langle\Phi_k|\hatw \mu_j|\Phi_r \rangle
$.  The $(E_r - E_k)\chi^j_{kr}(t|z)$ contribution arises from the $\hatw
H_1$ part of ${\wt H}(t)_{\rm w2}(z)$.  The notation $\chi^j(t|z)$ is used to
distinguish the approximate solutions of (\ref{eq4.5}) from the exact
$X^j(t|z)$.

Observe that the non-zero term $(E_r - E_k)\chi^j_{kr}(t|z)$ in
(\ref{eq4.5}) means that the $i\hbar$ factor in front of the time derivative
term can not cancel against a similar $\hbar$ multiplier on the right-hand
side.  In this sense (\ref{eq4.5}) is very different in its $\hbar$ analytic
structure than the Moyal equation of motion (\ref{eq1.4b}), where such
cancellation does occur.  This means that methods for approximately solving
(\ref{eq4.5}) are not close parallels of the standard semiclassical
expansions, (\ref{eq1.5b}) or (\ref{eq3.8a}).

The zeroth order reduction of (\ref{eq4.5}) results if one replaces the $*$
operation by ordinary multiplication to obtain the matrix ODE system
\begin{eqnarray} 
i\hbar {\PD{}t}
\chi^{0,j}_{kr}(t|z) &=& (E_r - E_k)\chi^{0,j}_{kr}(t|z) \nonumber\\
 & + & \sum_{s=1}^N
\bigg[\Big(\chi^{0,j}_{ks}(t|z)M_{sr}(t|z) 
 - M_{ks}(t|z)\chi^{0,j}_{sr}(t|z)\Big)\bigg]\,.
\label{eq4.6}
\end{eqnarray}
At this level of truncation one has recovered the  classical path
approximations to the line shape theory.  The associated correlation
function results from replacing $X^j(t|z)$ in (\ref{eq4.3}) with
$\chi^{0,j}(t|z)$ and restricting the $k$ sum to $N$ terms.  To further
clarify this point consider again the rigid rotor example, \cf (\ref{eq1.1b}).
Write the classical
flow generated by $h_2$ in terms of its coordinate and momentum parts:
$g(t|z)=(q(t|z),p(t|z))$.  The molecular interaction $M$ becomes
\begin{equation}
\fl M_{kr}(t|z) = \sum_{l=0}^\infty \frac{4\pi}{2l+1}
V^{(l)}_{12}(|q(-t|z)|)\sum_{m=-l}^{m=l} Y_{lm}^*(\hatw q(-t|z))
\langle\Phi_k|Y_{lm}(\hatw Q) |\Phi_r \rangle \,. \label{eq4.7}
\end{equation}
As is evident, the anisotropic interaction potential $V^{(l)}_{12}$ is
evaluated along the classical path $q(-t|z)$.
The states $\{|\Phi_k \rangle\}$ are
eigenstates of the $\H_1$ system angular momenta, $J^2$ and $J_z$.
This means that $\langle\Phi_k|Y_{lm}(\hatw Q) |\Phi_r \rangle$ has an
explicit evaluation in terms of Clebsch--Gordan coefficients.  Viewed as
a phase space function $M(t|z)$ is a time dependent $\H_1^{(N)}$ valued
symbol which is $\epsilon$ independent.

Now consider solutions of (\ref{eq4.5}) that include the leading
noncommutative $*$ effects.  We organize this family of approximations as
an asymptotic series in the small parameter $\epsilon$, which
describes the deformation of the $\sigma_2$--star product, \cf
(\ref{eq2.12b}), about conventional $\H_1$ operator multiplication,
\numparts\label{eq4.8}
\begin{equation} \chi^{j}_{kr}(t|z) =
\sum_{n=0}^{\infty}\epsilon^n\chi^{n,j}_{kr}(t|z) = \chi^{0,j}_{kr}(t|z) +
\epsilon\chi^{1,j}_{kr}(t|z) + O(\epsilon^2)\,.  \label{eq4.8a} \end{equation}
The leading term $\chi^{0,j}_{kr}(t|z)$ is the classical path approximation.
Placing (\ref{eq4.8a}) in (\ref{eq4.5}) and extracting the equation for
$\chi^{1,j}(t|z)$ gives 
\begin{eqnarray} 
\fl i\hbar {\PD{}t} \chi^{1,j}_{kr}(t|z)
= (E_r - E_k)\chi^{1,j}_{kr}(t|z) + \sum_{s=1}^N
\bigg[\Big(\chi^{1,j}_{ks}(t|z)M_{sr}(t|z)  -
M_{ks}(t|z)\chi^{1,j}_{sr}(t|z)\Big)   \nonumber \\
+ \frac{i\hbar}{2}J_{\alpha\beta}^{(2)}
\Big(\PD{}{z_\alpha}\chi^{0,j}_{ks}(t|z)\PD{}{z_\beta}M_{sr}(t|z) -
\PD{}{z_\alpha}M_{ks}(t|z)\PD{}{z_\beta}\chi^{0,j}_{sr}(t|z)\Big)\bigg]\,. &&
\label{eq4.8b}
\end{eqnarray}
Here the initial condition is $\chi^{1,j}_{kr}(0|z) = 0$.
The form of this equation for $\chi^{1,j}$ is a version of (\ref{eq4.6}) with
an inhomogeneous term added.  The $J^{(2)}_{\alpha\beta}$ term on the right of
(\ref{eq4.8b}) records, to leading order, the noncommutative nature of the
$*$ product.  The $z$ derivatives of the interaction $M$ may be
expressed via the Jacobi fields of the classical trajectories.  By the chain
rule one has
\begin{equation} \PD{}{z_\alpha}M_{sr}(t|z) = \langle
\Phi_s|\big[\grad_\gamma({\hatw H}_{12})_{\rm w2})\big](g(-t|z)\big)|
\Phi_r\rangle g_{\gamma;\alpha}(-t|z)\,. \label{eq4.8c}
\end{equation}
\endnumparts

The pattern one sees in (\ref{eq4.8b}) for the determination of
$\chi^{0,j}_{sr}$ and $\chi^{1,j}_{sr}$ continues to higher order.  Given
the values of $\{\chi^{l,j}_{kr}\}_{l=0}^{n-1}$ the ODE system
for $\chi^{n,j}_{kr}$
results from combining (\ref{eq4.5}), (\ref{eq4.8a}) with the $*$
expansion (2.10).
Normally, one would employ expansions 2) and 3) to
a common order.  Thus the $O(\epsilon^2)$  consistent calculation of
$\chi^{2,j}_{kr}$ requires the addition of the $\gamma^{(2)}(H_2;-t) \langle
\Phi_k| (\hatw {H}_{12})_{\rm w2}(z)|\Phi_r\rangle$ contribution to
$M_{kr}(t|z)$.

Evaluation of the integral (\ref{eq4.3}) is demanding because
$\chi^j_{kr}(t|z)$ needs to be numerically determined for each point $z$ in
the six dimensional phase space $T^*_2$.  However the number of integration
variables may be significantly reduced by using the spherical tensor structure
present in this problem.  The quantities $(\wt \BGamma(t)\hatw \mu_j)_{\rm
w2}(z)$ and $(e^{-\beta\hatw h_2})\wig(z)\hatw \mu_j$ are rank one tensors
whose contraction is a scalar.
One may reduce integration (\ref{eq4.3}) to three parameters by representing
$z=(q,p)$ by the three Euler angles and three rotational invariants
$|q|$,$|p|$ and $q\cdot p$.  The Euler angle integrals involve Wigner functions
${\cal D}_{M,M'}^J(\alpha,\beta,\gamma)$ and can  be done analytically.
As an example of this type of phase space tensor reduction, see \cite[Section
 III]{MCQ98}.

The method of approximating $X^j_{kr}(t|z)$ by the system of equations (4.9)
depends upon representing the $\sigma_2$--star product by its
leading $\epsilon$ terms. In the computation of observables one must
have $\epsilon=1$.  As discussed in Appendix A this is an asymptotic
derivative expansion.  Roughly speaking, it will succeed if the higher order
terms $B^{n}\prec X^j_{kr}(t),M_{sr}(t)\succ(z)$, $n\ge 2$ are small.

It is useful to consolidate the correlation function results into a single
statement.  Assuming that $\chi^{0,j}_{kr}$ and $\chi^{1,j}_{kr}$ are
solutions of (\ref{eq4.6}) and (\ref{eq4.8b}), respectively, the
$O(\epsilon^1)$ representation is
\begin{equation}\label{eq4.9}
\fl C(t) = 
\frac{1}{h^{3}}\sum_{k=1}^N \langle\Phi_1|\hatw \mu_j|\Phi_k \rangle
\int_{T_{2}^{\ast}} dz\;
\Big(\chi^{0,j}_{k1}(t|z)+ \chi^{1,j}_{k1}(t|z)\Big)
(e^{-\beta \hatw h_2})\wig(z)\,.
\end{equation}
In our view the appropriate test of success of the approximating methods
introduced here will be through numerical implementation and application to
specific molecular systems.

The correlation function formulas (\ref{eq4.3}) and (\ref{eq4.9}) assume that
the perturber could be treated as a point particle.  This restriction is
easy to relax, by enlarging the Hilbert space $\H_1$ to include both the
radiator and perturber internal degrees of freedom.  Likewise, the presumed
$t=0$ state of the system $\hatw{\rho}= |\Phi_1\rangle\langle\Phi_1 |\otimes
e^{-\beta\hatw{h}_2}$ can take a variety of alternate forms which allow a
superposition of molecular eigenstates  $\{|\Phi_j\rangle\}_1^N$
and the replacement of  $e^{-\beta\hatw{h}_2}$ by any function of
$\hatw{h}_2$.  The correlation function of interest here is based on the
dipole operator $\hatw \mu_j$.  However, the method of this section
continues to apply if $\hatw \mu_j$ is replaced by any
operator $\hatw A$ having a $\hbar$-regular\cite{Vor78,MF81,ROB87} mixed
symbol.

\section{Conclusions}

     The spectral profile of the intensity of an emission or absorption line
is the Fourier transform of the dipole autocorrelation function $C_N(t)$.  In
this paper we have introduced a mixed Weyl symbol formalism to represent the
dynamics needed to construct the single perturber 
$C(t)$.  Within this approach, the
radiator--perturber relative separation variables are characterized by the
phase space $T^*_2$, which serves as the support for the operator valued mixed
symbols. Expectation values and in particular $C(t)$ are realized,
\cf (\ref{eq4.3}), by a phase space average over traces of the product of
symbols.  The result is a fully quantum theory of  spectral line shapes.  In
the mixed symbol representation there is never any need to refer to perturber
wave packets.

     Furthermore, the symbol equation of motion (\ref{eq4.5})
for the time evolving dipole moment, $\chi_{kr}^j(t|z)$, admits a natural
semiclassical expansion which is based on the derivative expansion of the
Moyal bracket.  The mixed symbol formalism embeds within itself the
classical path approximation, which appears as the leading order of the
semiclassical truncation.  The subsequent corrections arise from the
noncommutative nature of the $*$ product for the mixed symbol.

The simplifications that result from the finite state coupling approximation
1) and the semiclassical expansions 2) and 3), give equations of motion that
are suitable for numerical solution.  To order $O(\epsilon^1)$ the
equations (\ref{eq4.6}) and (4.9)
are no more elaborate than
those previously used to numerically compute the atom--atom collision problem
within the Moyal formalism, \cf \cite{MCQ98}. The first term of (\ref{eq4.9})
is entirely equivalent to the dipole autocorrelation function written in the
classical path approximation and thus contains the same information as, for
example, equation $(4.2)$ in Griem's treatise \cite{Griem64}.  It
must be emphasized, however, that (\ref{eq4.9}) gives the first two terms of
a completely quantum mechanical expression for $C(t)$.  The established
success of the classical path approximation in accounting for observed line
shapes means that the $O(\epsilon^0)$ version of the theory has much of the
correct physics built into it.

For heavy perturbers the classical path approximation works well. The
diatomic molecular radiator--atomic perturber case has received much
attention; for example, HCl--Ar collisions were studied by Nielsen and Gordon
\cite{Niel73}.  More recently, Looney and Herman \cite{Loon87} made a
comparison of comprehensive calculations of the ${\rm N}_2$--broadened
rotational lines in the fundamental band of HCl with state-of-the-art
experimental data \cite{Pine87} and found excellent agreement.  Evidence that
the $O(\epsilon^1)$ correction may be sufficient in many cases for an
accurate quantum mechanical treatment comes from the work of Smith, Giraud
and Cooper \cite{Smith76} on CO--He.  Their classical path calculations for
this light perturber system agreed with close--coupling calculations to
within 10\%.

Transforming the general autocorrelation function in (\ref{eq4.9}) through
(\ref{eq1.0}) to give the profile $I(\omega)$ can be difficult.  The line
shape problem has two frequently employed limits, the impact approximation
which applies to line centers and the quasi--static approximation which
describes the far spectral wings \cite{All82}.  Various unified
theories have been devised to connect these two regimes.  Similar strategies
will have to be applied to (\ref{eq4.9}) and (\ref{eq1.0}) in order to
compute actual profiles accurately.

The mixed symbol method of computing $C(t)$ has an important flexibility.
The total isotropic intermolecular interaction energy is $v_2 +V_{12}^{(0)}$,
\ie the central potential can be arbitrarily divided between $v_2$ and
$V_{12}^{(0)}$.  It is customary, in the classical path approximation, to set $V_{12}^{(0)}=0$.  This means that the trajectories generated by
$h_2=h_{2,0} + v_2$ are consistent with the full central potential.  However,
one can make other choices.  If $v_2=0$, then $h_2=h_{2,0}$ (kinetic
energy) and the classical trajectories are constant velocity straight lines.
Here all the isotropic potential resides in $V_{12}^{(0)}$.  For $h_{2,0}$
evolution, the $O(\epsilon^0)$ formula (4.5) in 2) is exact, namely 
${\wt H}(t)_{\rm w2}(z)=\hatw H_1 + (\hatw H_{12})_{\rm w2}(g(-t|z))$.  
In this circumstance
all the semiclassical behavior comes from approximation 3). This approach
would improve upon the usual straight line trajectory technique often used to
treat the broadening of atomic lines by ionic perturbers \cite{Griem64}.
Another appealing option is to place all the attractive part of the
intermolecular interaction in $V_{12}^{(0)}$.  Now $v_2$ will be purely
repulsive and for this reason the $h_2$ flow will not have any unstable fixed
points.  In this way the unstable fixed point long time
breakdown\cite{MCQ98,CR97,BGP98} of the standard semiclassical approximation
is avoided.

We have introduced the mixed Weyl symbol formalism in order to obtain a
new and computationally viable full quantum version of line shape theory.
However the mixed symbol representation is also applicable to any composite
system having distinct quantum and semiclassical degrees of freedom.  The
scattering of spin dependent particles and the Coulomb excitation of an atom
or a nucleus are additional examples of such systems.



\ack

The authors would like to thank F. H. Molzahn for critically reading the
manuscript.  This research was supported in part by grants to TAO and GCT from
the Natural Sciences and Engineering Research Council of Canada.  MFK wishes
to thank the University of Manitoba for a post-doctoral fellowship.  

\appendix

\appendix
\section{Derivative Expansions}

The theory of pseudodifferential operators \cite{Hor80,MF81}
provides an analysis of the $*$ product and its associated derivative
expansions.  For $\C$-valued symbols $f,g$ on $T_2^*$ the $*$ product is
defined by the scalar analog of (\ref{eq2.9b}), namely
\begin{equation} \label{eqA1}
\fl f*g=(\pi\hbar)^{-2d_2} \int\int dz'\,dz''\, f(z+z')g(z+z'')
\exp[ 2i(z'\cdot J^{(2)} z'')/\hbar ]\, .
\end{equation}
In the rigorous approach there are two important questions: a) The convergence
of the integral (\ref{eqA1}) and the status of the Groenewold series (2.8)
as an asymptotic expansion of $f*g$; b) The properties of the
operator corresponding to a given symbol, in particular the symbol which is
the result of the $*$ product and its approximations.

The answer to these questions is effected by restricting the symbols to
special classes of functions.  A standard and important example is the class
$S^{m}$.  Take $(x,\xi)$ to be dimensionless versions of $(q,p)$, and denote
by $S^{m}$ a set of $C^\infty(\R^{d_2}\times\R^{d_2})$ functions with
estimate
\begin{equation}\label{eqA2} |\grad_{x}^{j}\grad_\xi^{k} f(x,\xi)|
\leq C_{jk} \langle \xi \rangle^{m-|j|}\,,\qquad\qquad \langle \xi \rangle =
\sqrt{1+|\xi|^2}
\end{equation}
valid for all multi-indices $j=(j_1,\dots,j_{d_2})$, $k=(k_1,\dots,k_{d_2})$
and some constant $C_{jk}$, uniformly for $x\in \R^{d_2}$.

Then the following statements \cite[Theorem (2.49)]{FOL89} hold: 1)   The map
$(f,g)\to f*g$ is continuous from $S^{m_1}\times S^{m_2} \to S^{m_1+m_2} $.
2)  For $f\in S^{m_1}$ and $g\in S^{m_2}$, the remainder $R_N$ in the
Groenewold series belongs to the class $S^{m_1+m_2-N}$ and so the expansion
(\ref{eq2.G1a}) becomes asymptotic. In other words,  $R_N=O({\langle \xi
\rangle}^{m_1+m_2 - N})$ and the error term vanishes as either
$N\rightarrow\infty$ while $\langle \xi \rangle>1$, or as $\langle \xi
\rangle\rightarrow\infty$ while $N > m_1+m_2$.  In this sense the derivative
series (\ref{eq2.G1a}) is a valid asymptotic expansion even when the small
scale parameter $\epsilon$ is fixed at unity.

\end{document}